\documentclass[final,authoryear,5p]{elsarticle}

\usepackage{amssymb}
\usepackage{graphicx}
\usepackage{multirow}
\usepackage{txfonts}
\usepackage{amssymb}

\usepackage[ps2pdf,%
a4paper=true,%
breaklinks=true,%
colorlinks=true,%
pdfauthor={First Author et al.},%
pdftitle={Template for manuscripts in Advances in Space Research}%
]{hyperref}

\journal{Advances in Space Research}

\begin{document}

\begin{frontmatter}

%\title{Evolution of the Temporal and Spectral properties in 2010 $\&$ 2011 outbursts of H~1743-322}
\title{Evolution of the temporal and the spectral properties in 2010 and 2011 outbursts of H~1743-322}

\author{Dipak Debnath\corref{cor}\fnref{footnote1}}
\address{Indian Centre For Space Physics, 43 Chalantika, Garia Station Road, Kolkata, 700084, India}
\cortext[cor]{Corresponding author}
\ead{dipak@csp.res.in}
\fntext[footnote1]{Tel.: + 91 33 24366003/24622153, Extn: 26; Fax: + 91 33 24366003/24622153, Extn: 28}

\author{Sandip K. Chakrabarti\fnref{footnote2}}
%\address{Indian Centre For Space Physics, 43 Chalantika, Garia Station Road, Kolkata, 700084, India}
\address{S. N. Bose National Center for Basic Sciences, JD-Block, Salt Lake, Kolkata, 700098, India}
\fntext[footnote2]{Also affiliated to Indian Centre For Space Physics, 43 Chalantika, Garia Station Road, 
Kolkata, 700084, India}
\ead{chakraba@bose.res.in}

%\author{Anuj Nandi\fnref{footnote2}}
\author{Anuj Nandi}
\address{Space Astronomy Group, SSIF/ISITE Campus, ISRO Satellite Centre, Outer Ring Road, Marathahalli,
Bangalore, 560037, India}
%\fntext[footnote2]{Also affiliated to Indian Centre For Space Physics, 43 Chalantika, Garia Station Road, 
%Kolkata, 700084, India}
\ead{anuj@isac.gov.in}

\date{Accepted; Received }

\begin{abstract}

The Galactic black hole candidate H~1743-322 exhibited two X-ray outbursts in rapid succession: 
one in August 2010 and the other in April 2011. We analyze archival data of this object from the 
PCA instrument on board RXTE (2-25 keV energy band) to study the evolution of its temporal and spectral 
characteristics during both the outbursts, and hence to understand the behavioral change of the accretion 
flow dynamics associated with the evolution of the various X-ray features. We study the evolution 
of QPO frequencies during the rising and the declining phases of both the outbursts. We successfully 
fit the variation of QPO frequency using the Propagating Oscillatory Shock (POS) model in each 
of the outbursts and obtain the accretion flow parameters such as the instantaneous shock locations, 
the shock velocity and the shock strength. Based on the degree of importance of the thermal (disk black 
body) and the non-thermal (power-law) components of the spectral fit and properties of the QPO (if 
present), the entire profiles of the 2010 and 2011 outbursts are subdivided into four different 
spectral states: hard, hard-intermediate, soft-intermediate and soft. We attempt to explain the nature 
of the outburst profile (i.e., hardness-intensity diagram) with two different types of mass accretion 
flow.

\end{abstract}


\end{frontmatter}

\parindent=0.5 cm

%%%%%%%%%%%%%%%%%%%%%%%%%%%%%%%%%%%%%%%%%%%%%%%%%%%%%%%%%%%%%%%%%%%%%%%%%%%%%

\section{Introduction}

Galactic transient black hole candidates (BHCs) are the most fascinating objects to study in 
X-ray domain since these sources exhibit evolutions in their timing and spectral 
properties during their outbursts. Several attempts \citep{MR03,Belloni05,RM06,DD08,Nandi12}
%(McClintock \& Remillard 2003; Belloni et al. 2005; Remillard \& McClintock 2006; Debnath et al. 
%2008; Nandi et al. 2012) 
were made for a thorough study on the temporal and spectral evolutions of the transient 
black hole (BH) binaries during their outbursts. Various spectral states were identified
during different phases of the outburst. In general, four basic spectral states ($Hard$, 
$Hard-Intermediate$, $Soft-Intermediate$, $Soft$) are observed during the outburst of a
transient BHC \citep{MR03,Belloni05,Nandi12}. %(McClintock \& Remillard 2003; Belloni et al. 2005; Nandi et al. 2012).
One can find detailed discussions about these spectral states and their transitions in the 
literature 
\citep{HomanB05,Belloni10c,Dunn10,Nandi12}. %(Homan \& Belloni 2005; Belloni 2010; Dunn et al. 2010; Nandi et al. 2012).
%It was reported by \citet{Nandi12} %Nandi et al. (2012) 
%that these four basic spectral states make a hysteresis-loop during the 
%2010-2011 outburst of GX 339-4, in the sequence: $hard \rightarrow hard-intermediate \rightarrow
%soft-intermediate \rightarrow soft \rightarrow soft-intermediate \rightarrow hard-intermediate
%\rightarrow hard$. 
It was also reported by several authors \citep{Fender04,HomanB05,Belloni10c,Nandi12} that the 
observed spectral states form a hysteresis loop during their outbursts.
Also, these different spectral states of the hysteresis-loop are found to be associated 
with different branches of a q-like plot of X-ray color vs intensity i.e., the hardness-intensity 
diagram (HID) \citep{MC03,HomanB05}. %(Maccarone \& Coppi 2003; Homan \& Belloni 2005a).

The transient low-mass Galactic X-ray binary H~1743-322 was first discovered \citep{Kaluzienski77} %Kaluzienski \& Holt 1977)
with the Ariel-V All-Sky Monitor and subsequently observed with the HEAO-1 satellite \citep{Doxsey77} %(Doxsey et al. 1997) 
in X-rays during the period of Aug-Sep, 1977. During the 1977-78 outburst, the
source was observed several times in the hard X-ray band of $12-180$ keV 
energy range with the HEAO-1 satellite \citep{Cooke84}. %(Cooke et al. 1984).
The observation revealed that the soft X-ray transient (based on the $1-10$ keV spectral 
properties) also emits X-rays in the energy range of $10-100$ keV \citep{Cooke84}. %(Cooke et al. 1984).
\citet{White84} %White \& Marshall (1984) 
categorized the source as a potential black hole candidate (BHC) based 
on the `color-color' diagram using the spectral data of the HEAO-1 satellite.

After almost two decades, in 2003, the INTEGRAL satellite discovered signatures of renewed activity 
%(though some X-ray activity was reported by EXOSAT in 1984 \citep{Reynolds99}) %(Reynolds et al. 1999)) 
in hard X-rays \citep{Revnivtsev03} %(Revnivtsev et al. 2003)
and later, RXTE also verified the presence of such an activity \citep{Markwardt03}. %(Markwardt \& Swank 2003).
During the 2003 outburst, the source was continuously and extensively monitored in X-rays 
\citep{Parmar03,Homan05,Remillard06,McClintock09}, %(Parmar et al. 2003; Homan et al. 2005b; Remillard et al. 2006),
IR \citep{Steeghs03}, %(Steeghs et al. 2003), 
and in Radio bands \citep{Rupen03} %(Rupen et al. 2003)
to reveal the multi-wavelength properties of the source. 
The multi-wavelength campaign on this source during its 2003 and 2009 outbursts were also carried 
out by \citet{McClintock09,Miller-Jones12} respectively.

The low-frequency as well as high frequency quasi-periodic oscillations (QPOs) along with a strong 
spectral variability are observed in the 2003 and other outbursts of the source in RXTE PCA data 
\citep{Capitanio05,Homan05,Remillard06,Kalemci06,Prat09,McClintock09,Stiele13}. 
%is another important discovery. 
These have resemblance with several other typical Galactic black hole candidates (e.g., GRO~J1655-40, 
XTE J1550-564, GX 339-4 etc.). Another important discovery of large-scale relativistic 
X-ray and radio jets associated with the 2003 outburst \citep{Rupen04,Corbel05} %(Rupen et al. 2004; Corbel et al. 2005)
put the source in the category of `micro-quasar'. This was also reconfirmed by \citet{McClintock09}, 
from their comparative study on the timing and the spectral properties of this source with XTE J1550-564.

Recently in 2010 and 2011, the transient black hole candidate H~1743-322 again exhibited outbursts 
\citep{Yamaoka10,Kuulkers11} %(Yamaoka et al. 2010; Kuulkers et al. 2011)
with similar characteristics of state transitions 
\citep{ShaposhnikovT10,Shaposhnikov10,Belloni10a,Belloni10b,Belloni11} 
%(Shaposhnikov \& Tomsick 2010a; Shaposhnikov 2010b; Belloni et al. 2010a,b, 2011) 
as observed in other outburst sources \citep{HomanB05,Nandi12}.
%\citep[][and references there in]{DD08,DD10}. (Debnath et al. 2008, 2010 and references there in).
Recently, \citet{Altamirano12} reported a new class of accretion state dependent $11$~mHz QPO 
frequency during the early initial phase of both the outbursts under study.

RXTE has observed both these outbursts on a daily basis, which continued for a time period of around two months. 
We made a detailed study on the temporal and the spectral properties of H 1743-322 during these two outbursts 
using archival data of PCA instrument on board RXTE satellite. Altogether $49$ observations starting 
from 2010 August 9 (MJD = 55417) to 2010 September 30 (MJD = 55469) of the 2010 outburst are analyzed in this paper.
%October 20 (MJD = 55489) are analyzed. 
After remaining in the quiescence state for around seven months, H~1743-322 again became active in X-rays 
on 2011 April 6 (MJD = 55657), as reported by \citet{Kuulkers11}. %Kuulkers et al. (2011).
RXTE started monitoring the source six days later (on 2011 April 12, MJD = 55663).
Here, we also analyze RXTE PCA archival data of $27$ observations spread over the entire outburst, 
starting from 2011 April 12 to 2011 May 19 (MJD = 55700). %June 8 (MJD = 55720).
The preliminary results of this work were already presented in COSPAR 2012 \citep{DD12}. %(Debnath et al. 2012)

Apart from the 2010 and 2011 outbursts, there are six outbursts of H~1743-322 observed
by RXTE in recent past. Detailed results of these outbursts have already been reported in the 
literature \citep{Capitanio09,McClintock09,Dunn10,Chen10,Coriat11,Miller-Jones12} 
%(Capitanio et al. 2009; Dunn et al. 2010; Chen et al. 2010)
and the evolution of all outbursts typically follow the `q-diagram' in the hardness-intensity plane 
\citep[see for example,][]{MC03,HomanB05}, 
%(see for example, Maccarone \& Coppi 2003; Homan \& Belloni 2005a),
except the 2008 outburst which does not follow the `standard' outburst
profile and is termed as the `failed-outburst' \citep{Capitanio09}. %(Capitanio et al. 2009).

Although the mass of the black hole has not yet been measured dynamically,
there are several attempts to measure the mass of the black hole based on the
timing and spectral properties of H~1743-322. 
From the model of high frequency QPOs based on the mass-angular momentum (i.e., spin of the 
black hole) relation, \citet{Petri08} %P\'{e}tri (2008)
predicted that the mass can fall in the range between $9 M_\odot$ to $13 M_\odot$.
%Here we assumed the mass of the black hole to be at $9 M_\odot$. 

The evolution of QPO frequency during the outburst phases of the transient BHCs has been 
well reported for a long time \citep{Belloni90,Belloni05,DD08,Nandi12}.
%During both the rising and the declining phases of these two outbursts, evolution of low frequency QPOs are observed. 
Same type of QPO evolutions were observed during both the rising and the declining phases of these two outbursts as 
of other black hole candidates, such as, 2005 outburst of GRO J1655-40 \citep{skc05,skc08}, %(Chakrabarti et al. 2005, 2008),
1998 outburst of XTE J1550-564 \citep{skc09} %(Chakrabarti et al. 2009)
and 2010-11 outburst of GX 339-4 \citep{DD10,Nandi12}. %(Debnath et al. 2010; Nandi et al. 2012).
The successful interpretation of these QPO evolutions with the Propagating Oscillatory Shock (POS) 
model \citep{skc05,skc08} %(Chakrabarti et al. 2008)
motivated us to fit the QPO evolutions of the recent outbursts of H~1743-322 with the same model. 
From the model fit, accretion flow parameters are calculated (see, Table 1 below).

This {\it Paper} is organized in the following way: In the next Section, we discuss about the 
observation and data analysis procedures using HEASARC's HEASoft software package. In \S 3, we present 
temporal and spectral results of our observation. 
In \S 3.1, the evolution of light curves (2-25 keV count rates) and hardness ratios of the 
2010 and 2011 outbursts of H~1743-322 are discussed.
In \S 3.2, we compare the evolution of the hardness-intensity diagrams of these two outbursts.
%2010 and 2011 outbursts of H~1743-322 
%in a single hardness-intensity diagram.
In \S 3.3, we show the time evolving (decreasing or increasing) nature of QPO frequency observed 
during rising and declining phases in both the outbursts (2010 and 2011) of 
H~1743-322 and apply the propagating oscillatory shock (POS) model
to explain the variations of the centroid QPO frequency over time. 
%We compare the results of our fits of these two outbursts during the rising 
%and the declining phases on a daily basis  and derive the flow properties (such as the locations 
%and strengths of the shock waves). We conclude that the oscillation of the shocked 
%accretion flow \citep{MSC96, Ryu97, skc04} %(Molteni et al. 1996, Ryu et al. 1997) 
%could be responsible for the generation and evolution of QPOs in this object as well.
In \S 3.4, we present the spectral analysis results and classify the entire duration of the 
outbursts into four spectral states: hard, hard-intermediate, soft-intermediate and soft. 
%We observe that the spectral states completes a cyclic order during the 
%both outbursts of H~1743-322, $hard \rightarrow hard-intermediate \rightarrow
%soft-intermediate \rightarrow soft \rightarrow soft-intermediate \rightarrow hard-intermediate
%\rightarrow hard$. 
Finally, in \S 4, we present the brief discussion and concluding remarks.

\section {Observation and Data Analysis}

The campaigns carried out with RXTE cover the entire 2010 and 2011 outbursts of 
H~1743-322 starting from 2010 August 9 (MJD = 55417) to 2010 September 30 (MJD = 55469) %October 20 (MJD = 55489) 
and from 2011 April 12 (MJD = 55663) to 2011 May 19 (MJD = 55700). %June 8 (MJD = 55720). 
We analyzed archival data of the RXTE PCA instrument 
and follow the standard data analysis techniques as done by \citet{Nandi12}. %Nandi et al. (2012).
The HEAsoft 6.11 version of the software package was used to analyze the PCA data. We extract data 
from the most stable and well calibrated proportional counter unit 2 (PCU2; all the three layers are
co-added).
%We used the standard RXTE data analysis software package HEAsoft 6.11. 

For the timing analysis, we use the PCA Event mode data with a maximum timing 
resolution of $125\mu s$. To generate the power-density spectra (PDS), we use the 
``powspec" task of XRONOS package with a normalization factor of `-2' to have the expected
`white' noise subtracted rms fractional variability on 2-15 keV (0-35 channels of PCU2) 
light curves of $0.01$ sec time bins. The power obtained has the unit of 
rms$^2$/Hz. %In general, low frequency QPOs ($<$ 20 Hz) are observed during the hard and intermediate spectral 
%states \citep{MR03,vdK04,DD08} %(McClintock \& Remillard 2003; van der Klis 2004; Debnath et al. 2008) 
%of outbursting BH candidates.
Observed QPOs are generally of Lorentzian type \citep{Nowak00,vdK05}. %(Nowak et al. 2000; van der Klis 2005).
So, to find centroid frequency of QPOs, power density spectra(PDS)  are fitted with Lorentzian profiles and fit error 
limits are obtained by using ``fit err" command. %task of the XRONOS package.
For the selection of QPOs in PDS, we use the standard method \citep[see][]{Nowak00,vdK05} based on the coherence parameter $Q$ 
(= $\nu$/$\Delta\nu$) and amplitudes (= \% rms), where $\nu$, $\Delta\nu$ are the centroid QPO frequency and 
full-width at half maximum respectively as discussed in \citet{DD08}. Here, for these two outbursts, 
observed $Q$ values and amplitudes are varied from $\sim 3-15$ and $\sim 5-16$ respectively.
In the entire PCA data analysis, we include the dead-time corrections and also PCA break down corrections 
(arising due to the leakage of propane layers of PCUs).

%If there is a variability in lightcurves, QPOs will be observed in PDS. 
%To find centroid frequency of these QPOs, PDS are fitted with Lorentzian profiles and 
%fit error limits are obtained by using ``fit err" task of the XRONOS package.
%While generating the PDS, we did not subtract the contribution due to 
%the Poissonian statistics. 

%Low frequency QPOs (LFQPOs) are generally observed during the hard and intermediate spectral 
%states \citep[][and references there in]{DD08} %(Debnath et al. 2008 and references there in) 
%of BHCs. During the present 2010 and 2011 outbursts of the BHC H~1743-322, LFQPOs are observed 
%during hard, hard-intermediate and soft-intermediate spectral states %\citep{DD11} %(Debnath et al. 2011) 
%of the rising and the declining outburst phases. 
%QPOs are generally of Lorentzian type \citep{Nowak00,vdK05}. %(Nowak et al. 2000, van der Klis 2005). 
%So, to find centroid frequency of QPOs, PDS are fitted with Lorentzian profiles and fit error 
%limits are obtained by using ``fit err" task of the XRONOS package.

For the spectral analysis, the standard data reduction procedure for extracting RXTE PCA 
(PCU2) spectral data are used. The HEASARC's software package XSPEC (version 12.5) is used 
for analyzing and modeling the spectral data. A fixed value of 1\% systematic error and the
hydrogen column density ($N_H$) of $1.6 \times 10^{22}$ \citep{Capitanio09} %(Capitanio et al. 2009)
for absorption model {\it wabs}, are used to fit the spectra. 
$2.5-25$ keV background subtracted PCA spectra are fitted with a combination of standard thermal (diskbb) 
and non-thermal (power-law) models or with only power-law component, where thermal photon contribution 
was much less (mainly in spectra from the hard and hard-intermediate spectral states). To achieve best 
fit, a single Gaussian Iron line $\sim 6.5$ keV is also used. The fluxes of different model components
of the spectra are calculated using {\it cflux} calculation method. 
 
\section {Results}

The accretion flow properties during the outburst phases of the transient BHCs can be understood 
in a better manner by studying X-ray properties of these sources both in temporal and spectral domains. 
It is pointed out by \citet{DD10} %Debnath et al. (2010) 
that depending upon the outburst light curve profiles, there are mainly two types of outbursting 
BHCs: one is `fast-rise slow-decay' (FRSD) type and the other is `slow-rise slow-decay' (SRSD) type. 
The source, H~1743-322 belongs to the first category. Although the general nature of the transient 
X-ray binaries is more complex \citep[see for example,][]{Chen97}. %(see for details, Chen et al. 1997).

\begin {figure}%[t]
%\vskip 0.45 cm
\vskip -0.25cm
\centering{
\includegraphics[scale=0.6,angle=0,width=6.5truecm,height=4.8truecm]{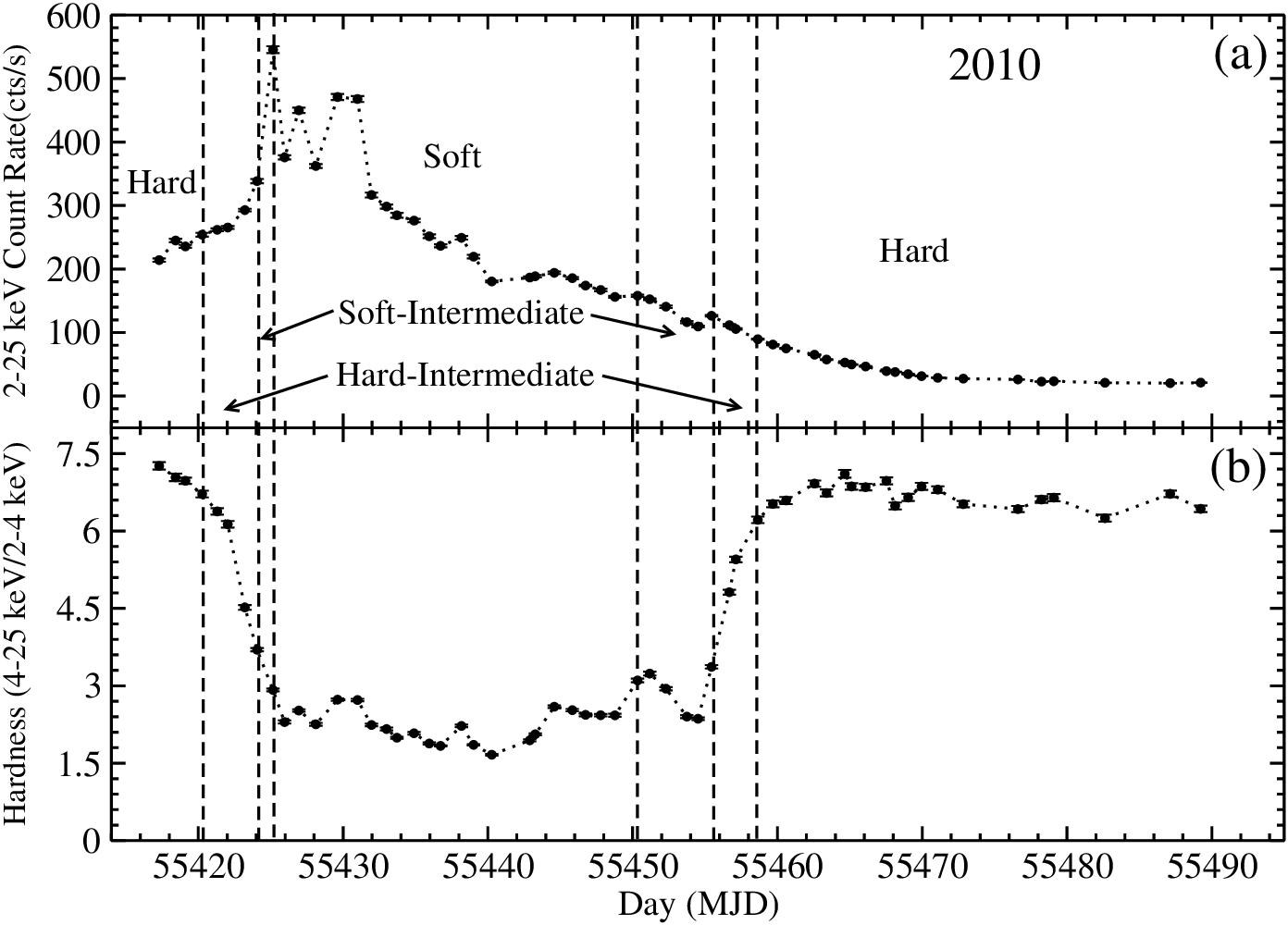}}%\hskip 0.4cm
\vskip 0.80cm
\centering{
\includegraphics[scale=0.6,angle=0,width=6.5truecm,height=4.8truecm]{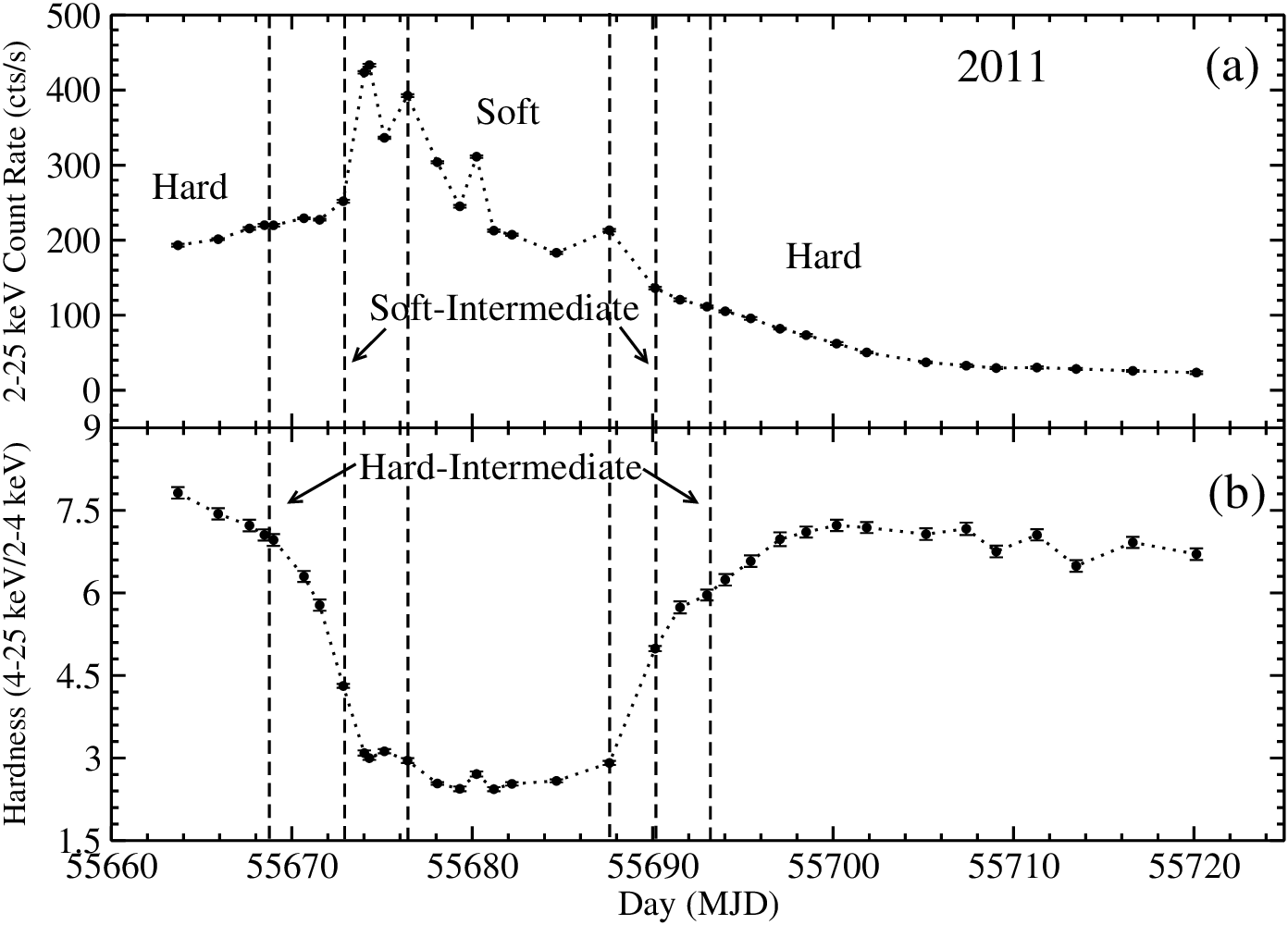}}
\caption{(a) 2-25 keV PCA light curves and (b) hardness ratios (4-25 keV versus 2-4 keV count ratio)
as a function of the MJD of 2010 (top panel) and 2011 (bottom panel) outbursts of H~1743-322 
are shown. The vertical dashed lines indicate the transitions between different spectral states. 
%(c) Hardness Intensity Diagram of the outbursts as observed with RXTE/PCA 
%are shown. Here $A-H$ and $a-h$ indicate start/state transitions/end of our observations from 2010 and 
%2011 outbursts respectively.
}
%\caption{Hardness Intensity Diagram of H~1743-322 during its 2010 (solid curve) and 2011 (dotted curve) 
%outbursts as observed with RXTE/PCA. The total count rates in $2-25$ keV energy band in along Y-axis 
%and the ratio of the count rates in the $4-25$ keV and $2-4$ keV bands is in the X-axis. 
%Here $A-H$ and $a-h$ indicate start/state transitions/end of our observation.
%X \& Y axes are in logarithmic scales. }
\label{kn : fig1}
\end {figure}

\subsection{Light curve evolution}

For studying X-ray intensity variations of the 2010 and 2011 outbursts of H~1743-322, we extract 
light curves from PCU2 data of RXTE/PCA instrument in different energy bands: $2-4$ keV ($0-8$ channels), 
$4-25$ keV ($9-58$ channels), and $2-25$ keV ($0-58$ channels). 
We have divided the $2-25$ keV energy band in the above two bands because $2-4$ keV photons mainly 
come from the thermally cool Keplerian disk, whereas the photons in the higher energy band ($4-25$ keV) 
come from the Comptonized sub-Keplerian disk (Compton corona). This fact may not be true always because 
the contributions for different spectral components also depend on accretion states. Variations of PCA 
count rates in $2-25$ keV energy band and hardness ratios between $4-25$ keV and $2-4$ keV count rates 
of the 2010 and 2011 outbursts of H~1743-322 are shown in Fig. 1(a-b). 

\subsection{Hardness-Intensity-Diagram (HID)}

In Fig. 2, we plot a combined $2-25$ keV PCA count rates of the 2010 and 2011 outbursts against X-ray 
color (PCA count ratio between $4-25$ keV and $2-4$ keV energy bands), which are well known as HID 
\citep{Fender04,HomanB05,DD08,Mandal10,Nandi12}. %(Homan \& Belloni 2005a; Debnath et al. 2008; Mandal \& Chakrabarti 2010).
%As mentioned in the previous section, depending upon the nature of the temporal and spectral 
%properties, we have classified the durations of the entire outbursts in four spectral states (hard, 
%hard-intermediate, soft-intermediate and soft), their state transitions are marked in the figure.
The marked points $A$, $B$, $C$, $D$, $E$, $F$, $G$, and $H$ are on MJD = $55417$, MJD = $55420$, 
MJD = $55424$, MJD = $55425$, MJD = $55450$, MJD = $55455$, MJD = $55458$, and MJD = $55469$ respectively 
for the 2010 outburst. Here points $A$ and $H$ respectively are the indicators of the start and the end of RXTE 
observations for the outburst and the points $B$, $C$, $D$, $E$, $F$, $G$ are the points 
on the days where the state transitions from hard $\rightarrow$ hard intermediate, hard-intermediate 
$\rightarrow$ soft-intermediate, 
soft-intermediate $\rightarrow$ soft, soft $\rightarrow$ soft-intermediate, soft-intermediate 
$\rightarrow$ hard-intermediate, and hard-intermediate $\rightarrow$ hard, respectively occurred. 
Similarly, the points $a$, $b$, $c$, $d$, $e$, $f$, $g$, and $h$ indicate MJD = $55663$, 
MJD = $55668$, MJD = $55672$, MJD = $55676$, MJD = $55687$, MJD = $55690$, MJD = $55693$, and 
MJD = $55700$ respectively for the 2011 outburst. 

\begin {figure}%[t]
\vskip 0.45 cm
\centering{
\includegraphics[scale=0.6,angle=0,width=6.5truecm,height=4.8truecm]{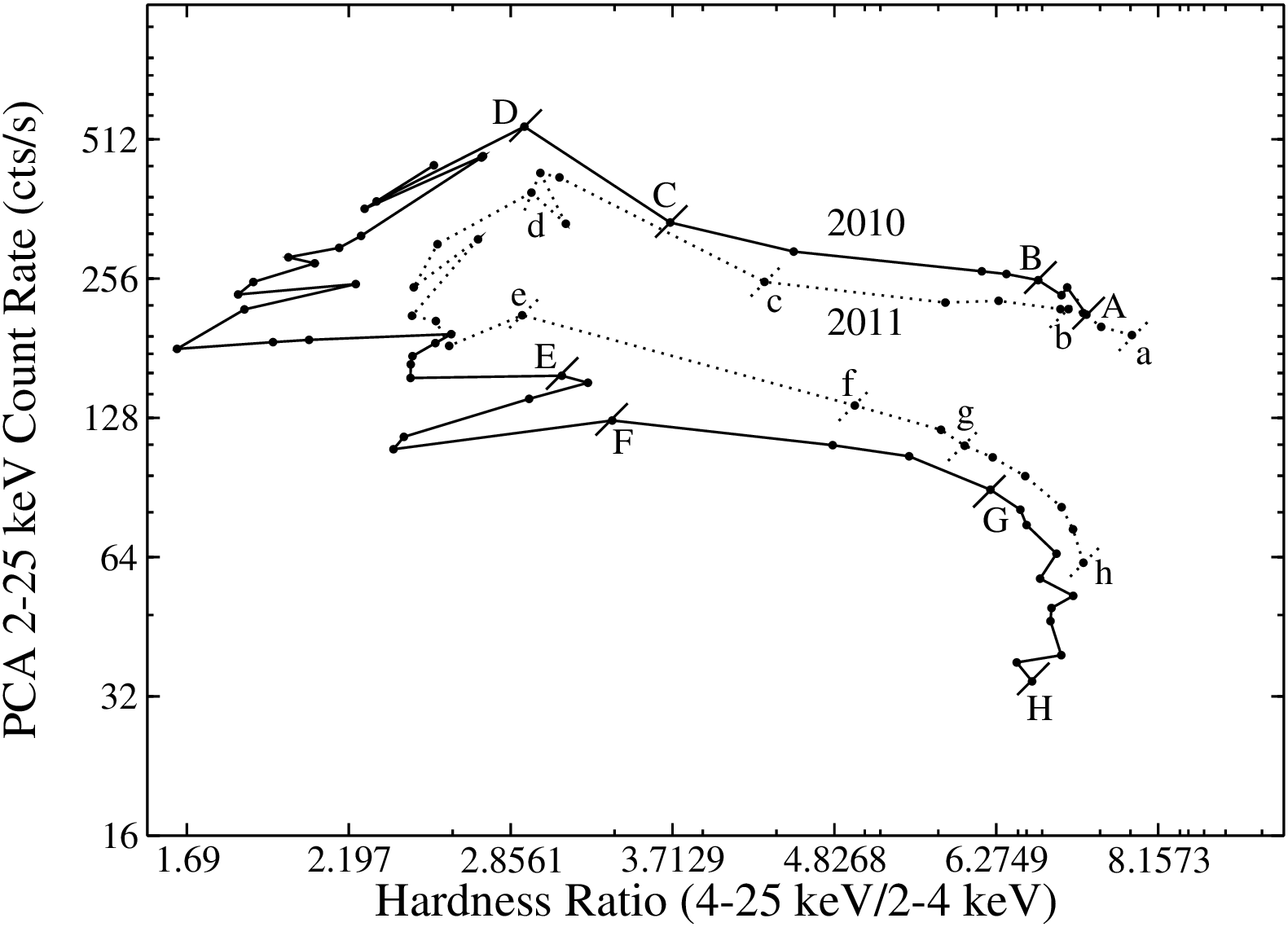}}
\caption{Hardness Intensity Diagram of the 2010 (solid curve) and 2011 (dotted curve) outbursts as 
observed with RXTE/PCA are shown. Here $A-H$ and $a-h$ indicate the start/state transitions/end of our 
observations from the 2010 and 2011 outbursts respectively.}
\label{kn : fig2}
\end {figure}

In the Figure, both the plots show similar nature and state transitions during the outburst, except 
that in the 2011 outburst, the PCA count rate is observed to be lower in rising phase and higher in 
declining phase of the outburst. During both the outbursts, the RXTE missed the initial rising days 
(supposed to be in the hard state) and the observational data was not available.

Recently, \citet{Altamirano12} also studied HIDs of both the outbursts. However, their analysis does not 
include spectral modeling of HIDs. From the detailed temporal and spectral study of these outbursts of 
H~1743-322, we have been able to connect different branches of the HIDs with different spectral states 
(see, Figs. 2, 6, 7). In the subsequent subsections, the variations of the spectral properties 
during the outbursts along with the POS model fitted evolutions of QPO frequency during the rising and 
the declining phases of the outbursts are discussed.

\subsection{Evolution of QPO frequency and its modeling by POS solution}

Studying temporal variability and finding QPOs in power density spectra (PDS) is an important aspect 
for any black hole candidate (BHC). It is observed (mainly at hard and hard-int-ermediate spectral states) 
that the frequency of QPOs are seen to evolve with time. LFQPOs are reported extensively in the 
literature, although there is some uncertainty about the origin of these QPOs. So far, many models are 
introduced to explain 
the origin of this important temporal feature of BHCs, such as trapped oscillations and disko-seismology 
\citep{Kato00}, %(Kato \& Manmoto, 2000; Ortega-Rodr\'iguez et al., 2002), 
oscillations of warped disks \citep{Shirakawa02}, %(Shirakawa \& Lai, 2002), 
accretion-ejection instability at the inner radius of the Keplerian disk \citep{Rodriguez02}, %(Rodriguez et al., 2002; Varniere et al., 2003), 
global disk oscillations \citep{Titarchuk00}, %(Titarchuk \& Osherovich, 2000),
and perturbations inside a Keplerian disk \citep{Trudolyubov99}, %(Trudolyubov, et al., 1999; Swank, 2001),
propagating mass accretion rate fluctuations in hotter inner disk flow \citep{Ingram11}, %(Ingram \& Done 2011),
and oscillations from a transition layer in between the disk and hot Comptonized flow \citep{Stiele13}. %(Stiele et al. 2013).
%In order to understand the origin of QPOs as well as the evolution of QPOs during outburts phases.
However, none of these models attempt to explain long duration continuous observations and the 
evolutions of QPOs during the outburst phases of transient BHCs. One satisfactory model namely shock 
oscillation model (SOM) by Chakrabarti and his collaborators \citep{MSC96}, shows that the oscillation 
of X-ray intensity could be due to the oscillation of the post-shock (Comptonizing) region. 
According to SOM, shock wave 
oscillates either because of resonance (where the cooling time scale of the flow is comparable to
the infall time scale; \citep{MSC96}) or because the Rankine-Hugoniot condition is not satisfied 
\citep{Ryu97} to form 
a steady shock. The QPO frequency is inversely proportional to the infall time ($t_{infall}$) in the 
post-shock region. The Propagating Oscillatory Shock (POS) model, which can successfully 
explain the evolutions of QPO frequency, is nothing but a special case (time varying form) of SOM.

As explained in our earlier papers on POS model \citep{skc05,skc08,skc09,DD10,Nandi12} during the 
rising phase, the 
shock moves towards the black hole and during the declining phase it moves away from the black hole. 
This movement of the shock wave depends on the non-satisfaction of Rankine-Hugoniot condition which 
is due to the temperature and energy differences between pre- and post- shock regions. 
%As because of that steady shock is not possible during the hard and hard-intermediate spectral states of both 
%rising and declining phases of the outbursts where QPO evolutions are observed, and shock is bound to move 
%inward or outward the black hole due to temperature, energy and density differences in pre- and post- shock regions. 
Moreover, sometimes in soft-intermediate states, QPOs are observed sporadically
\citep[for e.g., during the 2010-11 outburst of GX 339-4; see][]{Nandi12} and vanishes in soft spectral states 
and reappears in declining intermediate/hard states. This disappearance and appearance of QPO frequency 
depends on the compression ratio ($R$) due to the velocity/density difference in pre- and post- shock 
regions or could be due to the ejection of Jets \citep[see][]{RN13,Nandi13}.%(see Nandi et al. 2013). 
When $R = 1$, i.e., density of pre- and post- shock region more or less becomes the  same, a shock wave 
vanishes, and so does the QPO.

We now present the results of the evolution of QPO frequency observed in both rising and declining 
phases of both the outbursts. So far in the literature, there is no consensus
on the origin  of QPOs despite its long term discovery \citep{Belloni90,Belloni05}, 
other than our group \citep{skc05,skc08,skc09,DD10,Nandi12}. 
In this work, we have tried to connect the nature of the observed QPOs and their evolutions during the 
rising and the declining phases of the current outbursts with the same POS 
model and find their implications on accretion disk dynamics. From the fits, physical flow parameters, 
such 
as instantaneous location, velocity, and strengths of the propagating shock wave are extracted. 
Detailed modeling and comparative study between QPO evolutions observed in the rising and the 
declining phases of the outbursts of transient BHCs will be presented in our follow-up works, where we 
will compare the POS model fit parameters with the spectral/temporal properties (such as count rates, 
hardness ratios, spectral fluxes, photon indices etc.) of the BHCs. 
%Also in future, we have a plan to make a comparative study of the QPO frequency evolutions during all outbursts 
%of H~1743-322 or other multiple outburst sources observed with RXTE, if sufficient archival PCA data are during 
%the rising and the declining phases are available and will try 
This study can predict the mass of the BHCs, whose masses are not 
measured dynamically till now (for e.g., H 1743-322). Similarly, our study can predict the properties of QPOs
in subsequent days, once the data for the first few days is available.

The monotonically increasing nature of QPO frequency (from $0.919$~Hz to $4.796$~Hz for the 2010 
outburst and from $0.428$~Hz  to $3.614$~Hz for the 2011 outburst) during the rising phases and 
the monotonically decreasing nature of QPO frequency (from $6.417$~Hz to $0.079$~Hz for the 2010
outburst and from $2.936$~Hz  to $0.382$~Hz for the 2011 outburst) during the declining phases
of the recent successive two outbursts of H~1743-322  are very similar to what is observed in 
the 2005 outburst of GRO~J1655-40 \citep{skc05,skc08}, %(Chakrabarti et al. 2005, 2008),
1998 outburst of XTE~J1550-564 \citep{skc09}, %(Chakrabarti et al. 2009),
and 2010 outburst of GX~339-4 \citep{DD10,Nandi12}. %(Debnath et al. 2010; Nandi et al. 2012).
This motivated us to study and compare these evolutions with the same POS model solution. We found 
that during the rising and the declining phases of these two outbursts of H 1743-322, QPO evolutions 
also fit well with the POS model. The POS model fitted parameters (for e.g., shock location, strength, 
velocity etc.) are consistent with the QPO evolutions of GRO~J1655-40, XTE~J1550-564, and GX~339-4. 
%, with having fitted reduced $\chi^2 \sim 1$ for ising phases amd $\sim 2$ for delcining phases. 
The POS model fitted accretion flow parameters of the 2010 and 2011 outbursts of H~1743-322 are 
given in Table 1. Only noticeable difference observed during the present QPO frequency 
evolutions of H~1743-322 with that of the 2005 outburst of GRO~J1655-40 and 2010-11 outburst of 
GX~339-4 is that during both the rising phases of GRO~J1655-40 and GX~339-4 outbursts, the shock was 
found to move in with a constant speed of $\sim 2000~cm~s^{-1}$, and $\sim 1000~cm~s^{-1}$ respectively, 
whereas during the same phases of the current two outbursts of H~1743-322, the shock was found to move 
in with an acceleration. On the other hand, during the declining phase for all these outbursts 
of GRO~J1655-40, GX~339-4, and H~1743-322, the shock was found to be moved away with constant acceleration. 
It is also noticed that during both the rising and the declining phases of the 2010 outburst, the shock moved away 
with an acceleration twice as compared to that of 2011 outburst. It seems to be an 
interesting result, which may occur due to the lack of supply of matter (mostly Keplerian) into the disk 
from the companion that could have created a sudden `void' in the disk for the shock to move 
away rapidly outward.
%because it could provide the clues about the mechanism that determines the speed of the propagation of the shock wave.

According to the POS solution \citep{skc08,skc09,DD10,Nandi12}, %(Chakrabarti et al. 2008, 2009; Debnath et al. 2010; Nandi et al. 2012)
one can obtain the QPO frequency if one knows the instantaneous shock location 
or vise-versa %\citep[see Eqs. 1-3 of ][]{skc08} %(see, Eqs. 1-3 of Chakrabarti et al. (2008)) 
and the compression ratio ($R$ = $\rho_+$/$\rho_-$, where $\rho_+$ and $\rho_-$ are the densities 
in the post- and the pre- shock flows) at the shock. According to POS model in the presence of a 
shock \citep{CM00,skc08}, %(Chakrabarti \& Manickam 2000; Chakrabarti et al. 2008),
the infall time in the post-shock region is given by, 
$$
t_{infall}\sim  r_s/v \sim  R r_s(r_s-1)^{1/2} ,
\eqno{(1)}
$$
where, $r_s$ is the shock location in units of the Schwarzschild radius $r_g=2GM/c^2$, $v$ is the
velocity of propagating shock wave in $cm~s^{-1}$. 

The QPO frequency happens to be inversely proportional to the in-fall time scale from the post-shock 
region. According to the shock oscillation model \citep{MSC96}, %(Molteni et al. 1996),
oscillations of the X-ray intensity are generated due to the oscillation of the post-shock region. 
This is also the centrifugal pressure supported boundary layer (or, CENBOL) which behaves as a 
Compton cloud in the \citet{CT95} %Chakrabarti-Titarchuk (1995)
model of two component accretion flow (TCAF). According to the numerical simulations of the sub-Keplerian 
(low-angular momentum) accretion which includes the dynamical cooling \citep{Ryu97} %(Ryu et al. 1997)
or the thermal cooling \citep{MSC96,skc04}, %(Molteni et al. 1996; Chakrabarti et al. 2004),
the frequency of the shock oscillation is similar to the observed QPO frequency for BHCs. 
Thus, the instantaneous QPO frequency $\nu_{QPO}$ (in $s^{-1}$) is expected to be
$$
\nu_{QPO} = \nu_{s0}/t_{infall}= \nu_{s0}/[R r_s (r_s-1)^{1/2}]. 
\eqno{(2)}
$$
Here, $\nu_{s0}= c/r_g=c^3/2GM$ is the inverse of the light crossing time of the black hole
of mass $M$ in unit of $s^{-1}$ and $c$ is the velocity of light. In a drifting shock scenario,
$r_s=r_s(t)$ is the time-dependent shock location given by
$$
r_s(t)=r_{s0} \pm v_0 t/r_g ,
\eqno{(3)}
$$
where, $r_{s0}$ is the shock location at time $t = 0$ (first QPO observed day) and 
$v_0$ is the corresponding shock velocity in the laboratory frame. The `+' ve sign in the second term 
is to be used for an outgoing shock in the declining phase and the `-' ve sign is to be used for the 
in-falling shock in the rising phase.  When the velocity of the shock wave (as in the rising phase of the 2005 GRO J1655-40 outburst)
%\citep{skc08}) %Chakrabarti et al. 2008)
is constant, $v=v_0$. For the accelerating case (as in the rising and declining phases 
of the 2010 and 2011 outbursts of H 1743-322) $v$ is time-dependent and can be defined as 
$v(t)=v_0 + a t$, where $a$ is the acceleration of the shock front.

Since in the presence of cooling, the shock moves close to the black hole, at the rising phase of the 
outburst, where the cooling gradually increases due to rise of the Keplerian rate, the shock wave moves 
towards the black hole and thus the QPO frequency rises on a daily basis. The reverse is true in the 
declining phases. The POS model fitted results of the QPO evolutions during rising and declining phases 
of the 2010 and 2011 outbursts are presented in the following sub-sections.

\begin {figure}%[h]
\vskip 0.45 cm
\centering{
\includegraphics[scale=0.6,angle=0,width=8truecm]{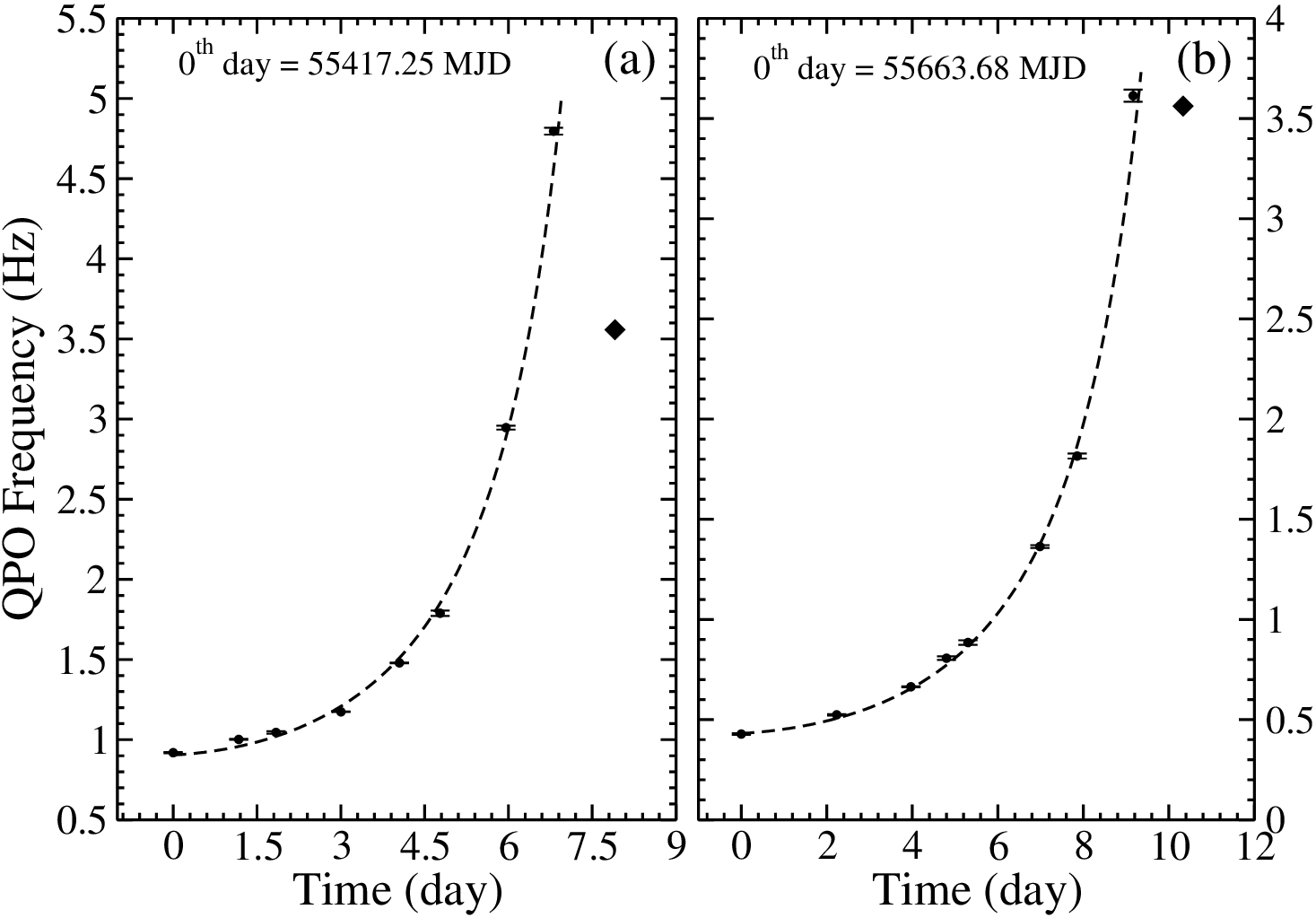}}
\caption{(a) Variations of the QPO frequency with time (in day) of the rising phase of the (a) 2010 outburst 
and (b) 2011 outburst which are fitted with the POS model solution (dashed curve). The diamond indicates the last 
event when the QPO was observed on (a) 2010 August 17 and (b) 2011 April 23, not included in the fits.}
\label{kn : fig3}
\end {figure}

\begin {figure}%[h]
\vskip 0.50 cm
\centering{
\includegraphics[scale=0.6,angle=0,width=8truecm]{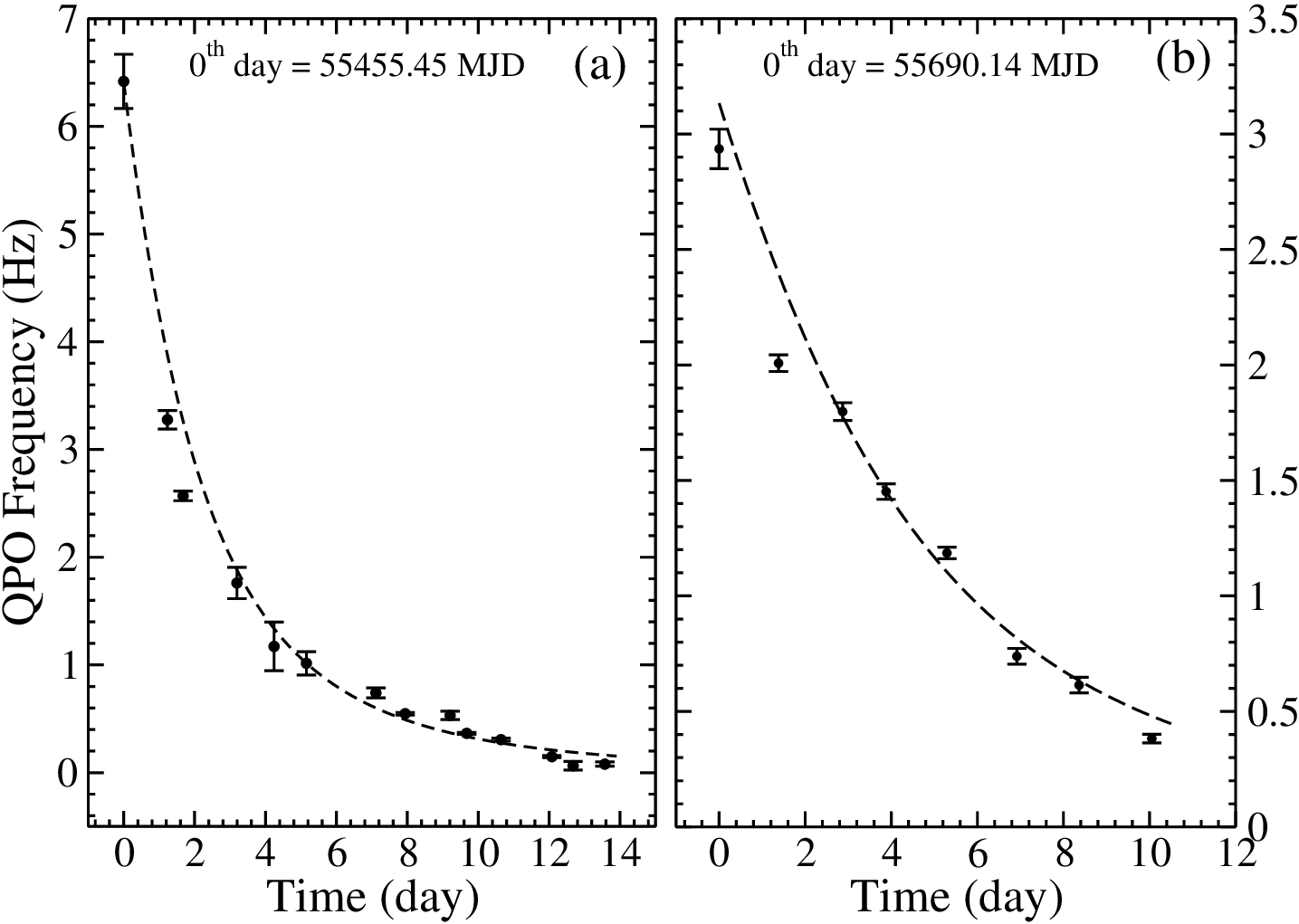}}
\caption{Variations of the QPO frequency with time (in day) of the declining phase of the (a) 2010 outburst 
and (b) 2011 outburst with the fitted POS model (dotted curve). Here, initial soft-intermediate state QPOs 
(as indicated by diamond points in Fig. 3) are not shown in the plot and also not included in the fit.}
\label{kn : fig4}
\end {figure}

\subsubsection{2010 QPO Evolutions}

%RXTE PCA archival data of $58$ observational IDs starting from 2010 August 9 
%(MJD = 55417) to 2010 October 20 (MJD = 55489). %The PCA light curves of $0.01$ sec bin for 
%the energy range of $2-15$ keV, are generated for all the observations and searched for the QPOs. 
The QPOs are observed in $23$ observations out of total $49$ observations starting from 
2010 August 9 (MJD = 55417) to 2010 September 30 (MJD = 55469) during the entire outburst. Out of these 
QPO observations, $9$ are observed in the rising phase and the remaining $14$ are observed in the 
declining phase of the outburst.

\vskip 0.1cm\noindent{$\bullet$ Rising Phase:}\vskip 0.1cm
On the very first observation day (2010 August 9, MJD = 55417), a QPO of type `C' \citep{vdK04} %(van der Klis 2004)
of $0.919$~Hz and its first harmonics of $1.842$~Hz were observed. On subsequent days, QPO frequencies 
are observed to be increased till 2010 August 16 (MJD = 55424, where $4.796$~Hz QPO is observed). 
From the next day, the frequency of the observed QPO (type `B') is decreased ($3.558$~Hz).
We have fitted this evolution of the QPO frequency with the POS model (Fig. 3a) and we found that 
the shock wave started moving towards the black hole from $\sim 428$ Schwarzschild radii ($r_g$) 
and reached at $\sim 181~r_g$ (Fig. 5a) within $\sim 7$~days. Also, we found that 
during this period, the shock velocity is varied from $\sim 180$~cm~s$^{-1}$ to $\sim 1133$~cm~s$^{-1}$ 
with an acceleration of $\sim 140$~cm~s$^{-1}$~d$^{-1}$ and the shock compression ratio $R$, 
which is inverse of the shock strength $\beta$, is changed from $1.39$ to $1.00$. Unlike 
2005 GRO~J1655-40 or 2010 GX~339-4, %(Debnath et al. 2010), 
%\citep{skc08} or 2010 GX~339-4 \citep{DD10},
we did not start with the strongest possible shock ($R = 4$) in the present case. 
This is because RXTE missed this object in the first few days of observation.
In the first 'observed' day, the shock has already moved in and 
the QPO frequency is already too high ($\sim 1$~Hz).
According to our model, if the RXTE monitoring started a few days earlier, we would have observed mHz QPOs
as in other black hole sources. 
The compression ratio $R$ decreased with time by the relation $1/R \rightarrow 1/R_0 + \alpha (t_d)^2$, 
where $R_0$ is the initial compression ratio (here $R_0 = 1.39$), $t_d$ is the time in days (assuming 
first observation day as 0$^{th}$ day). Here, $\alpha$ is a constant ($=0.0060$) which determines how 
the shock (strength) becomes weaker with time and reaches its lowest possible value when $R = 1$. In principle,
these parameters, including shock propagation velocity
can be determined from the shock formation theory when the exact amount of viscosity
and cooling effects are supplied \citep{C90}. %(Chakrabarti, 1990).

\vskip 0.1cm\noindent{$\bullet$ Declining Phase}\vskip 0.1cm
The source is seen to move to this phase on 2010 September 16 (MJD = 55455), when a QPO of $6.417$~Hz 
frequency is observed. On subsequent days, the observed frequency of the QPO decreases, and it reaches 
to its lowest detectable value of $79$~mHz on the 2010 September 30 (MJD = 55469) within a period of 
$\sim 13.6$~days. Before Sept. 16th, QPOs are sporadically observed at around $2 - 2.5$~Hz 
starting from 2010 September 11 (MJD = 55450). 
According to the POS model fit (shown in Fig. 4a), the shock is observed to recede back starting from 
$\sim 65~r_g$ till $\sim 751~r_g$ (Fig. 5a). The shock compression ratio $R$ appears to remain constant 
at $3.33$. Also, during this phase, the shock velocity varies from $\sim 560$~cm~s$^{-1}$ to 
$\sim 1578~cm~s^{-1}$ due to an acceleration of $75$~cm~~s$^{-1}$~d$^{-1}$.

\begin {figure}%[h]
\vskip 0.45 cm
\centering{
\includegraphics[scale=0.6,angle=0,width=8truecm]{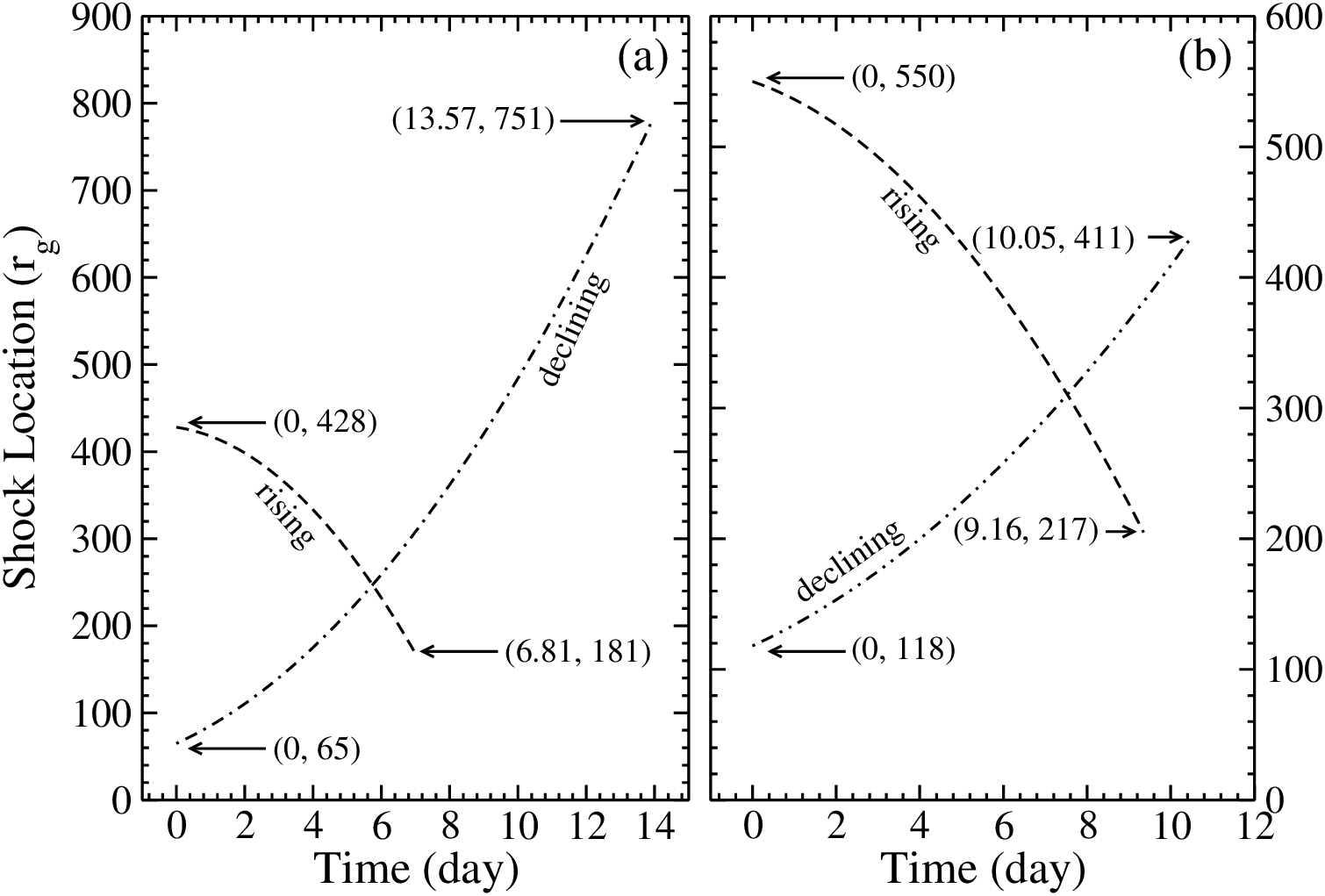}}
\caption{Variation of the shock locations (in $r_g$) during the rising and the declining 
phases of the (a) 2010 and (b) 2011 outbursts of BHC H~1743-322 (see text for details). } 
%During the rising phases of the outburst, the shock seems to be stagnated and oscillated at around 
%$170 - 190~r_g$ for the next few days and then disappears.}
\label{kn : fig5}
\end {figure}

\subsubsection{2011 QPO Evolutions}

%After remaining in quiescence state  for around seven months, H~1743-322 became X-ray active again 
%on 2011 April 6 (MJD = 55657), as reported by \citet{Kuulkers11}. %Kuulkers et al. (2011). 
%RXTE started monitoring the source six days later (on 2011 April 12, MJD = 55663), where $0.428$~Hz QPO is observed. 
%For the detailed study of the QPO properties, we analyze RXTE PCA archival data of 
%$35$ observational IDs spread over the entire  observed outburst, starting from 2011 April 12 
%to 2011 May 19 (MJD = 55700). For finding low and intermediate frequency QPOs, PCA light curves of 
%$2-15$ keV energy band and of $0.01$ sec time bin are generated for all the observations. 
The QPOs are observed in $19$ observations out of a total of $27$ observations spread over 
the entire outburst. Out of these $19$ observations, 
$11$ are observed in the rising phase and the remaining $8$ are in the declining phase of the outburst.
%$10$ are observed in the rising phase and the remaining $9$ are in the declining phase of the outburst.

\begin{table}%[h]
\addtolength{\tabcolsep}{-4.80pt}
%\scriptsize
\small
\centering
\caption{\label{table1} POS model fit of QPO Evolutions} 
\vskip 0.0cm
\begin{tabular}{|l|cccccccccc|}
\hline
Outburst&Period&$\nu_i$ & $\nu_f$ & $r_{si}$ & $r_{sf}$ & $v_i$ & $v_f$ & $a$ & $R_0$ & $\alpha$ \\
%\cline{2-9}
Phase  & (day)& (Hz) & (Hz) & ($r_g$) & ($r_g$) & (cm/s) & (cm/s) & (cm/s/d) &  & \\
%Phase  & (day)& (Hz) & (Hz) & ($r_g$) & ($r_g$) &  &  &  &  & \\
\hline
Ris.'10& 6.81&0.919&4.796&428&181&180&1133&140&1.39&0.0060 \\
Dec.'10&13.57&6.417&0.079& 65&751&560&1578& 75&3.33&...... \\
Ris.'11& 9.16&0.428&3.614&550&217&340&1137& 87&2.00&0.0055 \\
Dec.'11&10.05&2.936&0.382&118&411&460& 912& 45&2.78&...... \\
\hline
\end{tabular}
\leftline {Here, $\nu_i$ \& $\nu_f$ are the initial and the final POS model QPO frequencies}
\leftline {respectively. $r_{si}$ \& $r_{sf}$ are the initial and the final shock locations}
\leftline {respectively. $v_i$ \& $v_f$ \& $a$ are the initial and the final velocities}
\leftline {and the acceleration of the shock wave respectively. $R_0$ is} 
%\leftline {respectively. $v_i$ \& $v_f$ \& $a$ are the initial and the final velocities (in cm/sec)}
%\leftline {and the acceleration (in cm/sec/day) of the shock wave respectively. $R_0$ is} 
\leftline {the initial value of the shock compression ratio and $\alpha$ is a constant.}
%\leftline {Here, $\nu_i$ \& $\nu_f$ are the initial and the final POS model QPO frequencies respectively.}  
%\leftline {$r_{si}$ \& $r_{sf}$ are the initial and the final shock locations respectively. $v_i$ \& $v_f$ \& $a$ are}
%\leftline {the initial and the final velocities and the acceleration of the shock wave respectively.}
%\leftline {$R_0$ is the initial value of the shock compression ratio and $\alpha$ is a constant.}
\end{table}

\vskip 0.1cm\noindent{$\bullet$ Rising Phase}\vskip 0.1cm
During this phase of the outburst, a QPO of frequency $0.428$~Hz is observed on the 
first RXTE PCA observation day (2011 April 12, MJD = 55663). Similar to the rising phase of the 2010 
outburst, QPO frequencies are observed to be increasing with time and reached its maximum 
value of $3.614$~Hz (as observed by RXTE) on 2011 April 21 (MJD = 55672). On 2011 April 23 
(MJD =55674) and 2011 April 25 (MJD = 55676), the frequencies of the observed QPOs are seen to be
at $3.562$~Hz and $3.306$~Hz respectively. The evolutionary track of the QPO frequency is fitted 
with the POS model (Fig. 3b) with the method same as that used for 2010 data and here also it is 
found that the shock wave moved towards the black hole starting from the launching of shock location 
at $\sim 550~r_g$ (Fig. 5b). This reached at $\sim 217~r_g$ within a period of $\sim 9$~days. 
From the POS solution, it is observed that during the evolution period, the shock velocity 
is varied from $\sim 340$~cm~s$^{-1}$ to $\sim 1137$~cm~s$^{-1}$ with the effect of the 
acceleration of $\sim 87$~cm~s$^{-1}$~d$^{-1}$ and the shock compression ratio $R$ varies from 
$2.00$ to $1.038$. The compression ratio $R$ followed the same equation as the rising phase of 
the 2010 outburst with different constant values of $R_0 = 2.00$ and $\alpha = 0.0055$. This is 
primarily because RXTE started observing at different days after the onsets of
these two outbursts. It is difficult to predict the acceleration of the shock front without knowing 
how the matter is supplied at the outer boundary, these are treated as parameters in the present 
solution.

\vskip 0.1cm\noindent{$\bullet$ Declining Phase}\vskip 0.1cm
The source is observed to reach at this phase of the QPO evolution on the 2011 May 9 (MJD = 55690), 
where the QPO of frequency $2.936$~Hz is observed. Subsequently, as in the 2010 outburst, the 
frequency of the observed QPO decreased with time and reached to its lowest detectable value 
of $0.382$~Hz on 2011 May 19 (MJD = 55700). Three days prior to the start of this phase of QPO 
evolution, QPO of $2.215$~Hz is observed on 2011 May 6 (MJD = 55687). This behavior was also 
seen in the declining phase of the 2010 outburst. Here also we have fitted with the POS model 
solution (Fig. 4b) as the 2010 outburst and found that the shock moved away from the black hole 
with accelerating velocity and constant shock strength ($\beta \sim 0.36$ i.e., $R \sim 2.78$). 
During the outburst phase of $\sim 10$~days, the shock wave was found to move from $\sim 118~r_g$ 
to $\sim 411~r_g$ (Fig. 5b) with a change of velocity from $\sim 460$~cm~s$^{-1}$ to 
$\sim 912$~cm~s$^{-1}$ due to an acceleration of $45$~cm~s$^{-1}$~d$^{-1}$.

\subsection{Evolutions of Spectral States in 2010 and 2011 Outbursts}

In the previous Section, we showed that the QPO frequencies increased in the first few days and then 
decreased (declining phase) systematically in both the outbursts and indeed similar to the other 
outbursts studied by the same group. The movements of the shock location is related to the spectral 
evolution and thus it is worthwhile to check if the spectral evolution of H~1743-322 is also similar 
to those studied earlier. For studying the spectral properties, we fit the RXTE PCA spectra of $2.5-25$~keV 
energy band with the combination of the thermal (disk black body) and the non-thermal (power-law) components 
or with only non-thermal (power-law) component. To achieve the best fit, a single Gaussian line $\sim 6.5$ 
keV was used. 
We found that only a non-thermal power-law component is sufficient to fit the initial rising and final 
declining phases of the PCA spectra in $2.5-25$ keV energy range. A similar kind of the spectral behavior also 
observed in GX 339-4, as studied by \citet{MBH09}.
For all observations, we kept hydrogen column density ($N_H$) for absorption model {\it wabs} to be fixed 
at $1.6 \times 10^{22}$ \citep{Capitanio09}. %(Capitanio et al. 2009).

\begin{table}%[h]
\addtolength{\tabcolsep}{-4.50pt}
%\scriptsize
\small
\centering
%\caption{\label{table1} Spectral properties of H~1743-322 during 2010 and 2011 outbursts} 
\caption{\label{table1} Spectral Evolutions of H~1743-322 during the 2010 and 2011 outbursts.} 
\vskip 0.0cm
\begin{tabular}{|l|ccccccc|}
\hline
%Spectral& \multicolumn{6}{|c|}{2010 Outburst} \\
Spec.&Obs. Id&UT&$T_{in}(keV)$&$\Gamma$&DBB Flux& PL Flux&$\chi^2$/DOF \\
%Spectral  & Obs. Id & UT & $T_{in} (keV)$ & $\Gamma$ & DBB Flux & PL Flux &$\chi^2$/DOF \\
\cline{2-8}States& \multicolumn{7}{|c|}{2010 Outburst} \\
\hline
HS  &X-01-00&10/08&$---$&$1.67_{-0.05}^{+0.09}$&$---$&$2.81_{-0.09}^{+0.08}$&40.7/50 \\
HIMS&X-02-03&14/08&$---$&$1.87_{-0.03}^{+0.02}$&$---$&$3.62_{-0.11}^{+0.10}$&28.9/50 \\
SIMS&X-04-00&17/08&$1.03_{-0.01}^{+0.01}$&$2.25_{-0.01}^{+0.01}$&$2.29_{-0.06}^{+0.06}$&$4.33_{-0.05}^{+0.05}$&34.0/44 \\
SS  &X-11-00&31/08&$0.77_{-0.02}^{+0.02}$&$2.32_{-0.01}^{+0.01}$&$1.40_{-0.01}^{+0.01}$&$1.15_{-0.02}^{+0.02}$&41.9/44 \\
SIMS&X-20-01&12/09&$0.76_{-0.05}^{+0.05}$&$2.13_{-0.04}^{+0.02}$&$0.38_{-0.02}^{+0.01}$&$1.45_{-0.02}^{+0.02}$&43.1/44 \\
HIMS&X-23-01&18/09&$---$&$1.92_{-0.02}^{+0.02}$&$---$&$1.30_{-0.02}^{+0.02}$&45.4/50 \\
HS  &X-25-01&23/09&$---$&$1.74_{-0.02}^{+0.01}$&$---$&$0.67_{-0.01}^{+0.01}$&34.9/50 \\
%HS  &X-01-00&10/08&$1.59_{-0.05}^{+0.02}$&$1.61_{-0.03}^{+0.02}$&$0.14_{-0.01}^{+0.01}$&$2.26_{-0.11}^{+0.07}$&34.1/43 \\
%HIMS&X-02-03&14/08&$1.34_{-0.01}^{+0.01}$&$1.86_{-0.02}^{+0.01}$&$0.11_{-0.01}^{+0.01}$&$3.50_{-0.10}^{+0.10}$&29.2/44 \\
%SIMS&X-04-00&17/08&$1.03_{-0.01}^{+0.01}$&$2.25_{-0.01}^{+0.01}$&$2.29_{-0.06}^{+0.06}$&$4.33_{-0.05}^{+0.05}$&34.0/44 \\
%SS  &X-11-00&31/08&$0.77_{-0.02}^{+0.02}$&$2.32_{-0.01}^{+0.01}$&$1.40_{-0.01}^{+0.01}$&$1.15_{-0.02}^{+0.02}$&41.9/44 \\
%SIMS&X-20-01&12/09&$0.76_{-0.05}^{+0.05}$&$2.13_{-0.04}^{+0.02}$&$0.38_{-0.02}^{+0.01}$&$1.45_{-0.02}^{+0.02}$&43.1/44 \\
%HIMS&X-23-01&18/09&$1.21_{-0.01}^{+0.01}$&$1.85_{-0.04}^{+0.02}$&$0.07_{-0.01}^{+0.01}$&$1.23_{-0.02}^{+0.02}$&27.8/43 \\
%HS  &X-25-01&23/09&$1.67_{-0.05}^{+0.01}$&$1.55_{-0.02}^{+0.01}$&$0.05_{-0.01}^{+0.01}$&$0.62_{-0.01}^{+0.01}$&26.9/43 \\
\hline
\cline{2-8}& \multicolumn{7}{|c|}{2011 Outburst} \\
\hline
HS  &Y-02-00&16/04&$---$&$1.65_{-0.01}^{+0.01}$&$---$&$3.02_{-0.03}^{+0.03}$&50.5/50 \\
HIMS&Y-02-05&20/04&$---$&$1.89_{-0.01}^{+0.01}$&$---$&$2.78_{-0.07}^{+0.06}$&45.7/50  \\
SIMS&Y-03-00&23/04&$0.99_{-0.02}^{+0.02}$&$2.22_{-0.04}^{+0.03}$&$1.60_{-0.03}^{+0.02}$&$3.64_{-0.05}^{+0.04}$&30.8/44 \\
SS  &Y-04-00&30/04&$0.79_{-0.02}^{+0.01}$&$2.25_{-0.01}^{+0.01}$&$0.78_{-0.03}^{+0.02}$&$1.60_{-0.02}^{+0.02}$&44.2/44 \\
SIMS&Y-05-00&06/05&$0.79_{-0.02}^{+0.02}$&$2.21_{-0.02}^{+0.01}$&$0.64_{-0.02}^{+0.01}$&$1.91_{-0.03}^{+0.02}$&45.0/44 \\
HIMS&Y-05-02&10/05&$---$&$1.91_{-0.02}^{+0.01}$&$---$&$1.56_{-0.03}^{+0.03}$&53.9/50 \\
HS  &Y-06-02&16/05&$---$&$1.71_{-0.02}^{+0.01}$&$---$&$1.04_{-0.01}^{+0.01}$&32.6/50 \\ 
%HS  &Y-02-00&16/04&$1.60_{-0.02}^{+0.02}$&$1.58_{-0.02}^{+0.02}$&$0.12_{-0.02}^{+0.01}$&$2.90_{-0.03}^{+0.02}$&47.3/44 \\
%HIMS&Y-02-05&20/04&$1.32_{-0.02}^{+0.02}$&$1.82_{-0.02}^{+0.02}$&$0.13_{-0.02}^{+0.02}$&$2.65_{-0.07}^{+0.06}$&35.7/44  \\
%SIMS&Y-03-00&23/04&$0.99_{-0.02}^{+0.02}$&$2.22_{-0.04}^{+0.03}$&$1.60_{-0.03}^{+0.02}$&$3.64_{-0.05}^{+0.04}$&30.8/44 \\
%SS  &Y-04-00&30/04&$0.79_{-0.02}^{+0.01}$&$2.25_{-0.01}^{+0.01}$&$0.78_{-0.03}^{+0.02}$&$1.60_{-0.02}^{+0.02}$&44.2/44 \\
%SIMS&Y-05-00&06/05&$0.79_{-0.02}^{+0.02}$&$2.21_{-0.02}^{+0.01}$&$0.64_{-0.02}^{+0.01}$&$1.91_{-0.03}^{+0.02}$&45.0/44 \\
%HIMS&Y-05-02&10/05&$1.13_{-0.01}^{+0.01}$&$1.76_{-0.02}^{+0.01}$&$0.11_{-0.01}^{+0.01}$&$1.45_{-0.03}^{+0.02}$&31.9/43 \\
%HS  &Y-06-02&16/05&$1.46_{-0.02}^{+0.01}$&$1.64_{-0.01}^{+0.01}$&$0.03_{-0.01}^{+0.01}$&$1.00_{-0.01}^{+0.01}$&30.0/44 \\ 
\hline
\end{tabular}
\leftline {Here, $T_{in}$ \& $\Gamma$ represent the values of disk black body temperatures}
\leftline {and power-law photon indices respectively, and corresponding model}
\leftline {fluxes (in $10^{-9}~ergs~cm^{-2}~s^{-1}$) in $2.5-25$ keV energy range are enlisted}
\leftline {in DBB \& PL Flux columns. The errors are calculated with 90\%}
\leftline {confidence. Here, X=95360-14, Y=96425-01 and UT in dd/mm format.}
\end{table}

Based on the degree of importance of the disk black body and power-law components (according to fitted 
component value and their individual flux) and nature (shape, frequency, $Q$ value, rms\% etc.) of QPO 
(if present), the entire outburst periods of 2010 and 2011 are divided into four different spectral states: 
hard (HS), hard-intermediate (HIMS), soft-intermediate (SIMS) and soft (SS) 
(see, \citet{HomanB05} for the definitions of these basic spectral states).
Out of these four spectral states, the low frequency quasi-periodic oscillations (LFQPOs) are observed during 
hard, hard-intermediate and soft-intermediate spectral states while according to POS, the QPO evolutions 
are observed only during the hard and hard-intermediate spectral states. In soft-intermediate states, 
QPOs are observed sporadically. In general, observed QPOs during the hard and hard-intermediate spectral 
states are of `C' type \citep{vdK04} %(van der Klis 2004)
with Q-value $\geq 3$ and rms $\geq 10\%$ and during soft-intermediate spectral state are of `B' type 
with lesser Q and rms value. During both the outbursts, these four spectral states are observed in the 
same sequence and completed a hysteresis-type loop, with hard spectral state in both the start and the 
end phases while other three spectral states in between. It is to be noted that during the spectral 
evolution, the soft state is observed only once, during the mid-region of the outburst (see, Fig. 1(a-b), 
Fig. 2, Fig. 6, and Fig. 7).
%During both the outbursts, these four spectral states are observed in a similar sequence of 
%$hard \rightarrow hard-intermediate \rightarrow soft-intermediate \rightarrow soft \rightarrow 
%soft-intermediate \rightarrow hard-intermediate \rightarrow hard$ (see Fig. 5 \& 6). 
In Table 2, the model fitted values of the disk black body temperature ($T_{in}$ in keV) and 
power-law photon index ($\Gamma$) and their flux contribution to the spectra in $2.5 -25$ keV energy 
range for seven observations, selected from seven different spectral states of the 2010 and 2011 
outbursts are enlisted.

Daily variations of the model fitted parameters and their flux contribution in $2.5 - 25$ keV  
spectra of the 2010 and 2011 outbursts are plotted in Figs. 6 \& 7 respectively. 
The variations of the black body temperature ($T_{in}$), the power-law photon index ($\Gamma$) and their flux 
contributions in $2.5 - 25$ keV energy range are shown in these Figures. These variations justify the 
spectral classifications.
The Figures also show clearly that the evolutions of the spectral parameters and model fluxes are similar 
during the same spectral states of the two consecutive outbursts of H~1743-322. 

\begin {figure}[t]
\vskip 0.50 cm
%\hskip 0.5cm
\centering{
\includegraphics[scale=0.6,angle=0,width=9truecm]{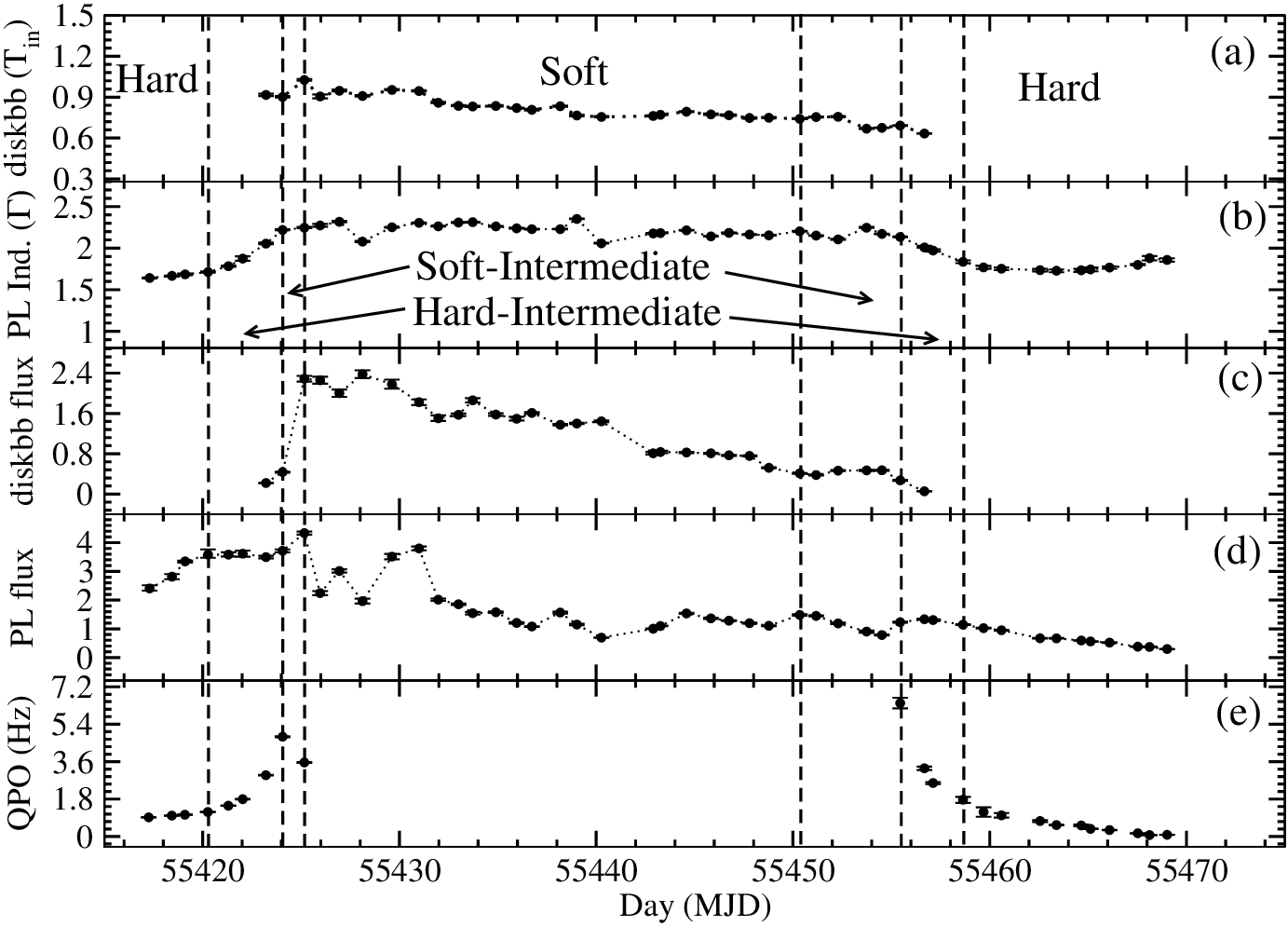}}
\caption{Variation of (a) disk black body temperature ($T_{in}$ in keV), (b) power-law photon index 
($\Gamma$), and $2.5 - 25$ keV fluxes (in $10^{-9}~ergs~cm^{-2}~s^{-1}$) of (c) diskbb, and 
(d) power-law models with day, observed in the 2010 outburst of H~1743-322, are shown. The 
parameters $T_{in}$ \& $\Gamma$ are obtained from the XSPEC model fit (using diskbb and power-law) 
of RXTE PCA spectra in $2.5-25$ keV energy band. In bottom panel (e), observed QPO frequency (in Hz) 
with day (MJD) are shown. The vertical dashed lines indicate spectral state transitions.} 
\label{kn : fig6}
\end {figure}

\subsubsection {2010 spectral evolution}

\noindent{\it (i) Rising Hard State: } 
Initial $\sim 3$ days of the RXTE observations (from MJD = 55417.3 to 55419.1) belong to this spectral 
state, where the spectra are fitted with only power-law (PL) component.
% along with an Iron line at $\sim 6.5$ keV.
So, during this phase, the spectra are dominated by the non-thermal photons without any signature of 
thermal photons.
%During this period PL photon index ($\Gamma$) and fluxes (in the range $2.5-25$ keV) 
%are varied in between $\sim 1.64$ to $\sim 1.68$ and $2.41$ to $3.35\times 10^{-9}~ergs~cm^{-2}~s^{-1}$ respectively.
The QPO frequency is observed to increase monotonically from $0.919$ Hz to $1.045$ Hz.

%%, according to two component advective flow (TCAF) model these photons mainy comes from sub-Keplerian disk. 
%%According to two component advective flow (TCAF) model, the reason for this type of evolutions in spectral parameters are 
%%due to more incoming Keplerian matter over sub-Keplerian matter. The flux variations shown in Fig. 5, 
%%justifies our interpetation, as because of disk black body (thermal) photons are mainly comes from Keplerian disk, 
%%where as power-law (non-thermal) photons are comes from sub-Keplerian disk.

\noindent{\it (ii) Rising Hard-Intermediate State: } 

In the following $\sim 5$ days (up to MJD = 55424.1), the source is observed to be at the hard-intermediate 
spectral state. Initial 3 days spectra are fitted without diskbb component, but in the rest of the two days spectra are 
fitted with the combination of diskbb (DBB) and power-law components. 
This is because as the day progresses, the spectra started becoming softer, due to enhanced supply 
of Keplerian matter. 
%During this phase, the observed $T_{in}$ are observed to be at $\sim 0.90$ keV and $\Gamma$ values are varied 
%from $1.71$ to $2.21$. 
During this state the spectra are mostly dominated by the non-thermal PL photons, although the thermal DBB 
rate is increased. 
%During this spectral state, $2.5-25$ keV DBB and PL model fluxes vary in the 
%range $0.22-0.44\times 10^{-9}~ergs~cm^{-2}~s^{-1}$ and $3.58-3.72\times 10^{-9}~ergs~cm^{-2}~s^{-1}$ respectively.
The QPO frequency is found to be increased monotonically from $1.045$ Hz to $4.796$ Hz.

\noindent{\it (iii) Rising Soft-Intermediate State: }
On the following day (MJD = 55425.2), the observed QPO frequency is decreased to $3.558$ Hz. After that 
no QPOs are observed for the next several days. We refer this particular observation as the soft-intermediate spectral state, because 
of sudden rise in DBB photon flux from its previous day value, % of $0.44\times 10^{-9}~ergs~cm^{-2}~s^{-1}$ 
%to $2.30\times 10^{-9}~ergs~cm^{-2}~s^{-1}$ in $2.5 - 25$ keV energy range. 
whereas the PL flux does not increase very much. 
%It is observed to be at $4.30\times 10^{-9}~ergs~cm^{-2}~s^{-1}$, 
%whereas the previous day value was at $3.70\times 10^{-9}~ergs~cm^{-2}~s^{-1}$ 
%in the same energy range. In this particular observation, the disk temperature ($T_{in}$) 
%and PL photon index ($\Gamma$) are observed at $1.03$ keV and $2.25$ values respectively.

\noindent{\it (iv) Soft State: }
The source is observed at this spectral state for the next $\sim 24$ days (up to MJD = 55448.8), where 
spectra are mostly dominated by thermal photons (i.e, low energy DBB photons). 
%During this phase, DBB and PL fluxes in $2.5 - 25$ keV energy range are varied from $2.30 \times 10^{-9}~ergs~cm^{-2}~s^{-1}$ to 
%$0.52 \times 10^{-9}~ergs~cm^{-2}~s^{-1}$ and from $2.20 \times 10^{-9}~ergs~cm^{-2}~s^{-1}$ to 
%$1.10 \times 10^{-9}~ergs~cm^{-2}~s^{-1}$ respectively. At the same time $T_{in}$ and $\Gamma$ values are 
%varied from $\sim 0.90$ to $\sim 0.70$ keV and from $\sim 2.20$ to $\sim 2.10$ respectively. 
No QPOs are observed during this spectral state (see Figs. 6 \& 7).

\noindent{\it (v) Declining Soft-Intermediate State: }
For the following $\sim 6$ days (up to MJD = 55454.5), the source is observed at this spectral state. Here,
$T_{in}$ and $\Gamma$ values are observed to be almost constant at $\sim 0.70$ keV and $\sim 2.20$ respectively. 
During this phase, disk black body flux is observed to be constant at $\sim 0.45 \times 10^{-9}~ergs~cm^{-2}~s^{-1}$, 
although there is an initial rise and then steady fall in the PL flux. 
%The PL flux is varied $1.50 - 0.80 \times 10^{-9}~ergs~cm^{-2}~s^{-1}$ during this short spectral period. 
Sporadic QPOs of $\sim 2$ Hz are observed during this spectral phase.

\noindent{\it (vi) Declining Hard-Intermediate State: }
The source is observed to be in this spectral state for the next $\sim 3.5$ days (up to MJD = 55457.1), 
where first two days spectra are fitted with combination of DBB and PL component and remaining
day's spectrum is fitted with only PL component.
%During this spectral period, $T_{in}$ and $\Gamma$ values are observed to be varied $0.69 - 0.63$ keV and $2.13 - 1.97$ respectively. 
The reason behind this is that as the day progresses, spectra became harder, because of lack 
of supply of Keplerian matter from the companion. 
%Observed DBB flux in $2.5 - 25$ keV energy range decreased sharply from $0.27 \times 10^{-9}~ergs~cm^{-2}~s^{-1}$ 
%to $0.06 \times 10^{-9}~ergs~cm^{-2}~s^{-1}$ within first two days, where as during this spectral period PL flux 
%is increased from $1.23 \times 10^{-9}~ergs~cm^{-2}~s^{-1}$ to $1.31 \times 10^{-9}~ergs~cm^{-2}~s^{-1}$.
It was also found that during this phase, the observed QPO frequency is monotonically decreased from $6.417$ Hz to $2.569$ Hz.

\noindent{\it (vii) Declining Hard State: }
This spectral state completes the hysteresis-like loop of the spectral state evolution (see Fig. 2). The source 
has been observed during this spectral state till the end of RXTE PCA observation of the 
2010 outburst. In this phase of evolution, the spectra are dominated by the non-thermal (power-law) flux. 
So, we fitted $2.5 - 25$ keV spectra with only PL model component.
% along with an Iron line of $\sim 6.5$ keV.
%During this period PL photon index ($\Gamma$) and fluxes (in the range $2.5-25$ keV) are varied in between 
%$1.72$ to $1.88$ and $1.14 - 0.29\times 10^{-9}~ergs~cm^{-2}~s^{-1}$ respectively.
Similar to the previous spectral state, the QPO frequency is found to be 
monotonically decreasing from $1.761$ Hz to $79$ mHz during this phase.

\begin {figure}[t]
\vskip 0.45 cm
\centering{
\includegraphics[scale=0.6,angle=0,width=9truecm]{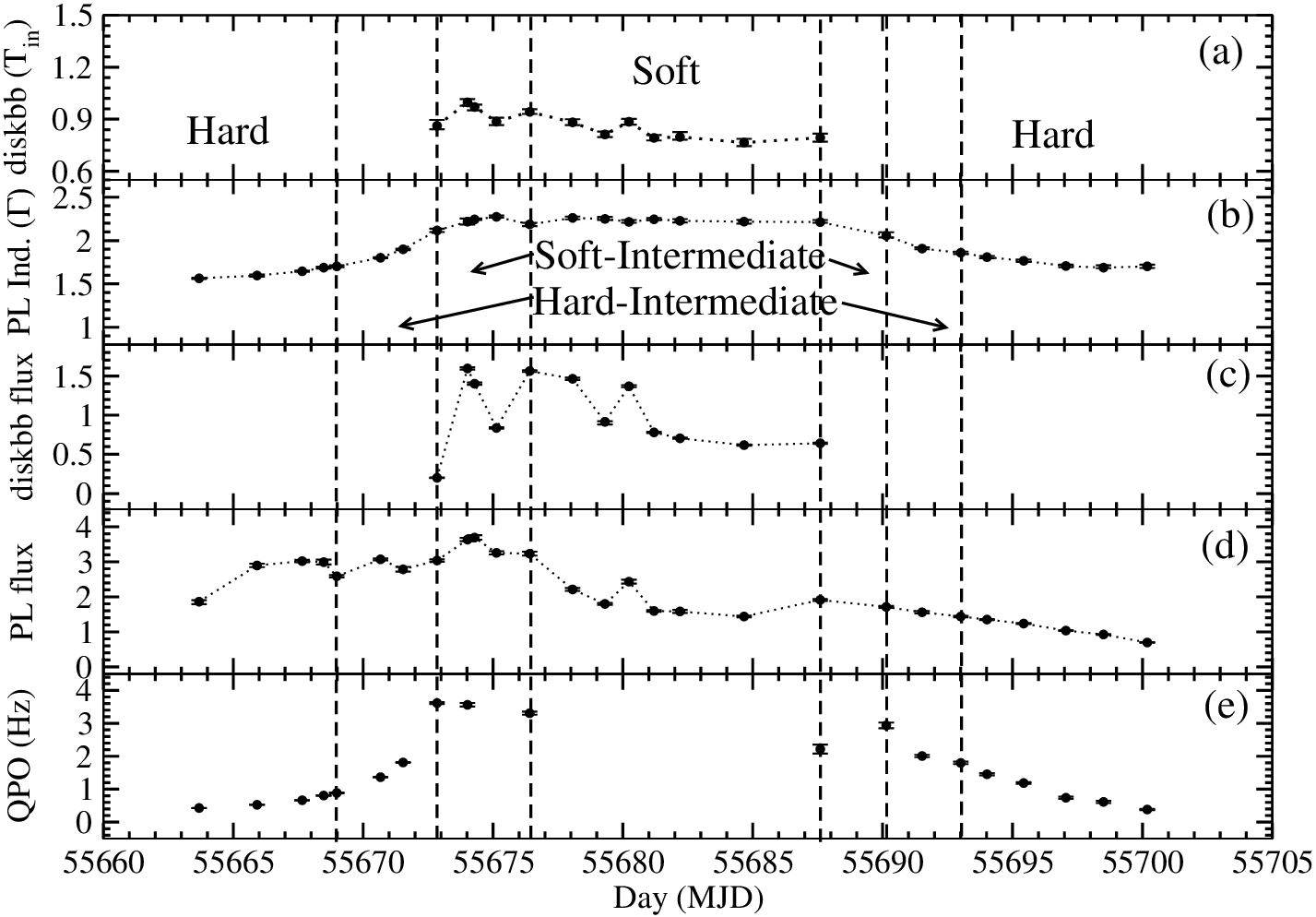}}
\caption{Same variations as in Fig. 6, except for the 2011 outburst of H~1743-322.} 
\label{kn : fig7}
\end {figure}

\subsubsection {2011 spectral evolution}

\noindent{\it (i) Rising Hard State:}
Initial $\sim 5$ days of PCA observations (from MJD = 55663.7 to 55668.5) belong to this spectral state, 
where spectra are fitted with only non-thermal power-law (PL) component.
% along with an Iron line at $\sim 6.5$ keV.
During this spectral state, the energy spectra ($2.5-25$ keV) are mostly dominated by non-thermal 
photons without any signatures of thermal photons.
%During this period PL photon index ($\Gamma$) and fluxes (in the range $2.5-25$ keV) are varied in between 
%$1.56$ to $1.70$ and $2.86\times 10^{-9}~ergs~cm^{-2}~s^{-1}$ to $3.02\times 10^{-9}~ergs~cm^{-2}~s^{-1}$ respectively. 
The observed QPO frequency is found to be monotonically increased from $0.428$ Hz to $0.807$ Hz.

\noindent{\it (ii) Rising Hard-Intermediate State:}
In the next $3$ observations (up to MJD = 55672.8), the source was observed to be in this spectral 
state, where first 2 days spectra are fitted without diskbb component, but remaining day's spectrum is 
fitted with the combination of diskbb (DBB) and power-law components. 
%During this spectral state, DBB temperature ($T_{in}$) is observed at $0.86$ keV and PL photon index ($\Gamma$) 
%is increased from $\sim 1.80$ to $\sim 2.12$. 
As the day progresses spectrum became softer, because of supply of more Keplerian matter (i.e, thermal
emission) from the companion. 
%During this period, in $2.5-25$ keV energy range DBB model flux is observed at 
%$0.20 \times 10^{-9}~ergs~cm^{-2}~s^{-1}$, where as at the same time, PL model flux in the same energy range 
%observed to be varied from $2.78$ to $3.07 \times 10^{-9}~ergs~cm^{-2}~s^{-1}$.
During this state, the QPO frequency is observed to be increased monotonically from $0.885$ Hz to $3.614$ Hz.

\noindent{\it (iii) Rising Soft-Intermediate State:}
The source is observed to be in this spectral state for the next $\sim 4.5$ days (up to MJD = 55676.4), 
where $T_{in}$ 
and $\Gamma$ values are observed to be almost constant at $\sim 0.90$ keV and $\sim 2.20$ respectively. A sharp rise 
in $2.5 - 25$ keV DBB flux over the previous state value is observed, where as the PL flux in the same 
energy range is observed to be nearly constant. 
%During this phase, $2.5 - 25$ keV DBB and PL model flux are varied 
%$0.80-1.60 \times 10^{-9}~ergs~cm^{-2}~s^{-1}$ and $3.70 -3.20 \times 10^{-9}~ergs~cm^{-2}~s^{-1}$ respectively.
As in the 2010 outburst, here also sporadic QPOs of frequency $\sim 3.5$ Hz are observed during this 
spectral state.

\noindent{\it (iv) Soft State:}
Next $\sim 8$ days (up to MJD = 55684.6), the source is observed to be in this spectral state, where $T_{in}$ 
and $\Gamma$ values are varied from $\sim 0.90$ to $\sim 0.80$ keV and from $\sim 2.30$ to $\sim 2.20$ respectively.
%Also during this spectral state, $2.5 - 25$ keV DBB and PL photon fluxes are varied 
%$1.50-0.70 \times 10^{-9}~ergs~cm^{-2}~s^{-1}$ and $2.20 -1.40 \times 10^{-9}~ergs~cm^{-2}~s^{-1}$ respectively. 
During this phase, the spectra are mostly dominated by low energy DBB flux (i.e., thermal emission) with 
decreasing in nature. 
%component fluxes are decreased slowly.
QPOs are not observed during this state, which are also missing during the soft state of the 2010 
outburst (see Figs. 6 \& 7).

\noindent{\it (v) Declining Soft-Intermediate State: }
On the next day (MJD = 55687.6), the source is observed to be in this spectral state with a weak
presence of thermal emission and the energy spectra started dominating by the PL flux. 
The particular observation showed a QPO signature at $2.215$ Hz. 
%is observed. In this observation, $T_{in}$ and $\Gamma$ values are observed as of our previous day 
%(MJD = 55684.6), soft state observation, at $0.79$ keV and $2.21$ respectively. 
%It also has been observed that, in this particular observation, DBB flux is almost as of previous 
%observation, but PL flux is increased from its previous observation.

%It also has been observed that, in this particular observation, $2.5-25$ keV DBB flux ($0.64\times 
%10^{-9}~ergs~cm^{-2}~s^{-1}$) is almost as of previous observation, but PL flux ($1.91\times 
%10^{-9}~ergs~cm^{-2}~s^{-1}$) is increased from its previous observation.

\noindent{\it (vi) Declining Hard-Intermediate State: }
After that up to MJD = 55691.5, the source was observed to be at this spectral state, where spectra are 
fitted without diskbb component. The spectra are dominated by non-thermal PL photons, because of 
lack of supply of Keplerian matter. 
%During this period PL photon index ($\Gamma$) and fluxes (in the range $2.5-25$ keV) are varied in 
%between $2.06$ to $1.91$ and $1.71 - 1.56\times 10^{-9}~ergs~cm^{-2}~s^{-1}$ respectively. 
QPOs are also observed during this spectral state and found to be decreased monotonically 
from $2.94$ Hz to $2.01$ Hz.

\noindent{\it (vii) Declining Hard State: }
At the final phase of the outburst, the source is found to be in the hard state again, which completes the 
hysteresis-like loop of the spectral state evolutions (see Fig. 2). Similar to the `canonical' 
hard state in the rising phase, here we also found that diskbb component is not essential to fit the PCA spectra in 
$2.5-25$ keV range, only PL component is sufficient to fit the spectra along with an Gaussian line at $\sim 6.5$ keV.
%During this spectral state, PL photon index and fluxes in the range $2.5-25$ keV) are observed to be varied 
%in between $1.86$ to $1.70$ and $1.44 - 0.69\times 10^{-9}~ergs~cm^{-2}~s^{-1}$ respectively. 
At the same time, during this spectral state, the QPO frequency is found to be decreased monotonically 
from $1.798$ Hz to $0.382$ Hz.

\section{Discussions and concluding remarks}

We carried out the temporal and the spectral analysis of the data of the 2010 and 
2011 outbursts of the black hole candidate H~1743-322. We studied the evolution of
quasi-periodic oscillation frequency during the rising as well as the declining phases.
We also studied the evolution of spectral states during both the outbursts. 
The variations of QPO frequencies can be fitted %very well 
assuming that an oscillating shock wave 
progressively moves towards the black hole during the rising phase and moves away from the 
black hole in the declining phase. Fundamentally, it is possible that a sudden rise in viscosity 
not only causes the Keplerian rate to rise but also causes the inner edge to move towards the 
black hole. Initially, the higher angular momentum flow forms the shock far away, but as the viscosity
transports the angular momentum, the shock moves in, especially so due to enhanced 
cooling effects in the post-shock region. The Keplerian disk moves in along with the shock.

This scenario accomplishes all that we observe in an outbursting source: 
(a) The QPO frequency rises/decreases with time in the rising/declining phase, mainly observed during the 
hard and hard-intermediate spectral states and during the soft-intermediate spectral state QPOs are seen 
sporadically \citep[see][]{Nandi13}. It is to be noted that shocks exist only in these states.
(b) The spectrum softens as the Keplerian disk moves in with a higher rate. (c) At the 
intermediate state(s), the Keplerian and the sub-Keplerian rates are similar. (d) During the 
declining phase, when the viscosity is reduced, the shock and the Keplerian disk 
moves back to a larger distance and the QPO 
frequency is also reduced. (e) The outflows can form only from the post-shock region (CENBOL), 
namely, the subsonic region between the shock and the inner sonic point. In softer states, the CENBOL 
disappears and the outflows also disappear. Our model predicts that since the QPOs could be due to the 
oscillation of the shocks, whose frequency is roughly the inverse of the infall time scale, the frequency 
gives the location of the shock when the compression ratio is provided. In our scenario, a strong shock ($R \sim 4$) starts 
at $\sim 1000 r_g$, but by the time it comes closer to the black hole, it becomes weaker due to the rapid cooling
by enhanced Keplerian disk rate. 
QPO ceases to exist when the compression ratio is unity. These constraints allowed us to compute the 
shock strength as a function of time.  

As far as the evolution of the spectral states during the two outbursts of the transient BHC H~1743-322 
is concerned, this can be well understood by the detailed study of the spectral properties. During 
both the outbursts, it has been observed that the source starts from the hard state and finally return back 
to hard state again after passing through the hard-intermediate, soft-intermediate and soft spectral states. 
It completes hysteresis loop of $hard \rightarrow hard-intermediate \rightarrow soft-intermediate \rightarrow 
soft \rightarrow soft-intermediate \rightarrow hard-intermediate \rightarrow hard$. 
Several attempts have already been made to understand these type of hysteresis spectral state transitions 
in black hole sources and to find their correlations with HIDs \citep{Meyer07,Meyer-Hofmeister09}, but one can 
easily explain this type of evolution of spectral states with the TCAF model \\ 
\citep{CT95}, %(Chakrabart \& Titatchuk 1995), 
where the low-angular momentum sub-Keplerian matter flows in nearly free-fall time scale, while the high angular 
momentum Keplerian matter flows in the slow viscous time scale \citep{Mandal10}. Initially the spectra are dominated by 
the sub-Keplerian flow and as a result, the spectra are hard. As the day progresses, more and more 
sub-Keplerian matter is converted to Keplerian matter (through viscous transport of angular momentum) and 
the spectra become softer, progressively through hard-intermediate (Keplerian rate slightly less than 
the sub-Keplerian rate), soft-intermediate (Keplerian rate comparable to the sub-Keplerian rate) and soft 
state (dominating Keplerian rate). When viscosity is turned off at the outer edge, the declining phase 
begins. At the declining phase of the outburst, the Keplerian rate starts decreasing, and the spectra start 
to become harder again. However, the spectrum need not be retrace itself, since the information about 
the decrease of viscosity had to arrive at the viscous time scale. This causes a hysteresis effect. But 
the spectra still follows the declining soft-intermediate, hard-intermediate and hard states.

%In general, LFQPOs can be observed during hard, hard-intermediate and soft-intermediate spectral states, while 
%according to the propagating oscillatory shock (POS) model, we can conclude that the evolutions of QPO frequency during 
%the rising and declining phases of the outbursts will be observed during the hard and hard-intermediate spectral states 
%only, while in the soft-intermediate state, QPOs can be observed sporadically on and off. 
In this work, we successfully applied the POS model fit evolutions of QPO frequency during both the rising and 
declining phases of two (2010 and 2011) outbursts of H 1743-322 and shock wave parameters related to the 
evolutions are extracted. Earlier, the same POS model was also applied to explain the evolution of QPO 
frequency of other black hole candidates (e.g., GRO~J1655-40, XTE~J1550-564, GX~339-4, etc.) very 
successfully \citep{skc05,skc08,skc09,DD10,Nandi12}. 
%(Chakrabarti et al. 2005, 2008, 2009; Debnath et al. 2010; Nandi et al. 2012).
All these objects seem to exhibit a similar behaviour as far as the QPO and spectral evolutions are concerned. 
In future, we will carry out detailed modeling and comparative study between QPO evolutions observed in 
other outbursts of H~1743-322 and other transient BHCs with this POS model and hence to understand accretion 
flow behaviours during the outburst phases more precisely.
However, the basic questions still remain: (a) What are the sources of enhanced viscosity? 
(b) Does it scale with the mass of the black hole or the mass of the donor? (c) Is the duration of 
the high viscosity phase (i.e., the duration between the end of the rising phase and the beginning of 
the declining phase) predictable, or it is totally random and depends mostly on the physical conditions 
of the donor? (d) Which processes decide the total time interval for which an outburst may last? And 
finally, (e) What determines the interval between two outbursts? If the cause is the enhancement of viscosity, 
then clearly it may be also random. We are in the process of exploring these aspects through comparison of 
all the known candidates. Recently, we have been able to include TCAF model in XSPEC as a local additive 
model, and from the spectral fit using this model directly we obtain instantaneous location of the 
shock ($r_s$) and compression ratio ($R$) other than two component (Keplerian and sub-Keplerian) accretion rates 
\citep[see][]{DD13a}. As we know from the POS model, one can determine the QPO frequency if the values of $r_s$ 
and $R$ are known or vise-versa (see, Eqn. 2). So, from the spectral fit, we will be able to predict the observed QPO 
frequency. The preliminary result on this work is already presented in a Conference Proceeding \citep{DD13b}.
%on {\it Recent Trends in the Study of Compact Objects: Theory and Observation} at Guwahati, Assam, India \citep{DD13}}.

%\section{Acknowledgments}

\end{document}